\documentclass[useAMS,usenatbib,twocolumn]{mn2e}
\usepackage{natbib}
\usepackage{times}
\usepackage{graphicx,amssymb,amsmath,fancybox,boxedminipage}
\usepackage{psfrag}
\usepackage{pstricks}
\usepackage{placeins}
\usepackage{sidecap}

\usepackage{booktabs}
\usepackage{ctable}
\usepackage{float}
\usepackage{multirow}
\usepackage{dcolumn}

\def\vmax{\ifmmode {\rm v_{max}}\else $\rm v_{max}$\fi}
\renewcommand{\H}[1]{\ifmmode x_{\rm H_{#1}}\else $x_{\rm H_{#1}}$\fi}
\newcommand{\Hp}[1]{\ifmmode x_{\rm H_{#1}^+}\else $x_{\rm H_{#1}^+}$\fi}
\newcommand{\e}{\ifmmode x_{\rm e} \else $x_{\rm e}$ \fi}
\newcommand{\z}{\ifmmode \zeta/n \else $\zeta/n$ \fi}
\newcommand{\zz}{\ifmmode \zeta_{-16}/n_3 \else $\zeta_{-16}/n_3$ \fi}
\newcommand{\Ot}{\ifmmode x_{\rm O_2} \else $x_{\rm O_2}$\fi}
\newcommand{\OH}{\ifmmode x_{\rm OH} \else $x_{\rm OH}$ \fi}
\newcommand{\CO}{\ifmmode x_{\rm CO} \else $x_{\rm CO}$ \fi}
\newcommand{\Op}{\ifmmode x_{\rm O^+}\else $x_{\rm O^+}$\fi}
\newcommand{\C}{\ifmmode x_{\rm C}\else $x_{\rm C}$ \fi}
\newcommand{\Cp}{\ifmmode x_{\rm C^+}\else $x_{\rm C^+}$\fi}
\def\Ht{\ifmmode {\rm H_2} \else H$_2$ \fi}

\title[CO/H$_2$, C/CO, OH/CO, and OH/O$_2$ in Dense Interstellar Gas]
{\vspace*{-0.9 truecm} CO/H$_2$, C/CO, OH/CO, and OH/O$_2$ in Dense Interstellar Gas: \\
From High Ionization to Low Metallicity \vspace*{-0.6 truecm}}

\author[S. Bialy \& A. Sternberg]
{Shmuel Bialy$^{\star1}$ and Amiel Sternberg$^1$
 \\
$^1$Raymond and Beverly Sackler School of Physics \& Astronomy, Tel Aviv University, Ramat Aviv 69978, Israel\\
 $^{\star}$ shmuelbi@mail.tau.ac.il\\
\vspace*{-0.5 truecm}
}

\date{Accepted: 2015 April 15.  Received: 2015 March 30; in original form: 2014 September 23 \vspace*{-0.5 truecm}}
\begin{document}

\maketitle
\begin{abstract}
\noindent
We present numerical computations and analytic scaling relations for
interstellar ion-molecule gas phase chemistry down to very low metallicities ($ 10^{-3} \times$ solar), and/or up to high driving ionization rates.  Relevant environments
include the cool interstellar medium (ISM) in low-metallicity dwarf galaxies,
early enriched clouds at the reionization and Pop-II star formation era, and in dense
cold gas exposed to intense X-ray or cosmic-ray sources.
We focus on the behavior for H$_2$, CO, CH, OH, H$_2$O and O$_2$,
at gas temperatures $\sim 100$~K, characteristic of a cooled ISM at low metallicities.
We consider shielded or partially shielded one-zone gas parcels, and solve 
the gas phase chemical rate equations for the
steady-state ``metal-molecule" abundances for a wide range 
of ionization parameters, $\zeta/n$, and metallicties, $Z'$. We find that the
OH abundances are always maximal near the H-to-H$_2$
conversion points, and that large OH abundances persist at very low metallicities
even when the hydrogen is predominantly atomic. We study the OH/O$_2$, C/CO and OH/CO abundance ratios, from large to small, as functions 
of $\zeta/n$ and $Z'$. Much of the cold dense ISM for the Pop-II generation 
may have been OH-dominated and atomic rather than CO-dominated and molecular.

\end{abstract}


\section{Introduction}
\label{sec: intro}

In this paper we present a numerical and analytic study
of interstellar gas-phase ion-molecule chemistry, from Galactic
conditions into the relatively 
unexplored domain of interstellar media at very low metallicities, 
and related chemical properties at 
high ionization rates. Relevant astrophysical environments include
the cool ISM in low metallicity dwarf galaxies, early enriched clouds
at reionization and the Pop-II star formation epoch, and in dense
cold gas exposed to intense Xray or cosmic-ray sources.
We consider the behavior for
a wide range of initiating hydrogen ionization rates
and gas densities. Our focus is mainly on the molecular cold-cloud
chemistry for H$_2$, CO, CH, OH, H$_2$O and O$_2$, and
the relative steady-state abundances of these species,
for varying metallicities and ionization parameters. Our discussion
includes a detailed analysis of the hydrogen-carbon-oxygen
gas-phase chemical networks for cold gas
in the transition to the low-metallicity limit.

There are several motivations for this work. First, 
observations of Population-II stars in the Galactic halo 
\citep{Beers2005}
have revealed stars with heavy element abundances orders of magnitude smaller than 
in the Sun and the interstellar medium (ISM) of the star-forming disk.
In the primitive halo star SDSS J102915+192927, for example,
an overall metallicity of less than $10^{-4}\times$solar has been reported
\citep{Caffau2011}. In many metal-poor stars the carbon/iron ratios are significantly enhanced (e.g.,~\citealt{Norris2013, Yong2013a, Carollo2014}) but even so the absolute carbon and oxygen abundances are very low, down to $10^{-3}$ compared to solar 
photospheric values. 
The very existence of these stars is evidence for interstellar media 
with metallicities far lower than in present-day galaxies.
Here we ask, and considering gas-phase ion-molecule 
formation-destruction chemistry 
in a cold dense ISM
how do the atomic and molecular constituents
for the heavy elements, especially carbon and oxygen, depend on
the overall metallicity, from solar down to very subsolar?

Star formation, and the processes that led to the
metal poor Pop-II stars, probably included
molecule formation to a level that enhanced gas cooling
and regulated the Jeans masses.
In pristine or in very low metallicity environments, even partial conversion to
H$_2$ enables efficient rotational line cooling 
in addition to atomic and ionic metal fine-structure emissions 
\citep{Bromm2003b, Santoro2006, Glover2007, Omukai2010, Glover2014}.
At high metallicities such as in Milky Way star-forming regions, 
complete conversion to H$_2$ occurs \citep[e.g.,][]{Sternberg2014} enabling the associated production of heavy molecules
\citep{Graedel1982, vanDishoeck1998, Tielens2013}.
For the cold ``dense-gas" components of the ISM,
what are the most abundant molecules containing heavy elements? 
For normal Galactic conditions at solar metallicity this is CO,
especially in regions shielded from photodissociating radiation.
Is this true also for a metallicity of 10$^{-3}$?

Very low metallicity stars have also been detected in 
nearby dwarf-spheroidal (dSph) galaxies \citep{Tafelmeyer2010}, e.g., 
most recently in the Galaxy satellite Segue 1 \citep{Frebel2014}. 
These dwarf galaxies may be directly related to the low-metallicity
damped Ly$\alpha$ absorbers (DLAs) observed from low to high redshifts
(e.g., \citealt{Pettini2008, Penprase2010, Rafelski2012}).
The low DLA metallicities, e.g.~down to $\sim 2\times 10^{-3}$ 
in the ($z$=5.1) absorber toward the QSO J1202+3235 \citep{Rafelski2012},
and the similar elemental abundance patterns, suggest that
the DLAs may represent the initial conditions
for the low-metallicity Pop-II stars and dSphs. 
The relative abundance patterns, including $\alpha$-element enhancements,
appear consistent with nucleosynthetic production in massive (initially metal-free) Population-III stars at the reionization epoch. Then, if ultra-low metallicity molecular clouds existed at reionization and in the protogalaxies in which the Pop-II stars formed,
what were their chemical properties?

Atomic and molecular carbon/oxygen chemical networks have been 
included in models for the cooling and fragmentation properties of
gravitationally contracting clumps and protostar formation at very low-metallicities
\citep[e.g.,][]{Omukai2000, Schneider2002, Omukai2005, Schneider2006, Jappsen2009B, Jappsen2009A, Dopcke2011, Klessen2012, Omukai2012, Chiaki2013}. In these models the chemistry is intrinsically time-dependent as molecules form
via gas-phase sequences in collapsing clouds.
The behavior is thus similar to molecule reformation zones behind dissociative shock waves
\citep[e.g.,][]{Hollenbach1989, Neufeld1989}.
In contrast, the focus of our paper is on the chemical properties
down to low metallicties
{\it of ionization-driven systems},
for which well defined steady-states exist.
This may be more appropriate for the chemical behavior of the 
bulk cold interstellar medium.
Such ISM components may be virialized and not immediately collapsing, or not even
gravitationally bound.
 As for the ISM in present-day galaxies we assume a (quasi)
heating-cooling equilibrium.
Thus our study is complementary to time-dependent models of the
kind presented by \citet{Omukai2005} for which a steady-state
is not defined.


A second motivation is the large range  
in ionization rates that may be expected for the 
neutral atomic/molecular ISM in galaxies
from low to high redshift and with varying star-formation rate.
For the standard neutral Galactic ISM, the ion-molecule chemistry and the production of ``heavy molecules" is driven by (low-energy) cosmic ray ionization. The ionization rate, as probed by observations of H$_3^+$, HCO$^+$, OH, OH$^+$, or H$_2$O$^+$, in diffuse and dense gas, and in the
intercloud medium (e.g., \citealt{VanDishoeck1986, VanderTak2000, Neufeld2010, Indriolo2012})
may vary with location in the Galaxy and with individual cloud thickness, 
but overall it seems to lie within the fairly narrow 
range $\sim 10^{-16}$ to 10$^{-15}$~s$^{-1}$ 
\citep{Dalgarno2006, Indriolo2012}. 
In as much as cosmic ray production depends on
star-formation and the subsequent
supernova explosions and shock particle accelerations, the internal
ionization rates may be significantly larger in more rapidly
star-forming and/or compact galaxies, compared to
3~M$_\odot$~yr$^{-1}$ within $\sim 10$~kpc for
the Milky Way. At high redshift and along the upper end of the 
main-sequence for star-forming galaxies \citep{Whitaker2012}, or for merging star-bursting systems, the surface star-formation rates may be orders
of magnitude larger than for the Galaxy (e.g.,~\citealt{ForsterSchreiber2009}; \citealt{Tacconi2013}), and this may imply 
correspondingly very high cosmic-ray fluxes \citep{Papadopoulos2010, Bayet2011, Mashian2013}.
The galaxy mass-metallicity relation \citep{Tremonti2004, Mannucci2010}
may then also imply a correlation between metallicity and global ionization rate
for the neutral ISM in galaxies.

The ionization rates may also be much larger than the
characteristic Galactic value in localized environments
exposed to X-rays, or enhanced cosmic-ray fluxes. Penetrating  X-rays and cosmic-rays are largely equivalent in their chemical effects, since the hydrogen ionization is always due to the secondary electrons produced by either primary X-ray photoionization 
or cosmic-ray impact ionization \citep{Lepp1996a, Maloney1996, Meijerink2005}. 
X-ray irradiation may play a dominating role in 
the inner envelopes and disks around young stellar objects
\citep{Igea1999, Stauber2005, Vasyunin2008}. 
External to the Galaxy, X-ray driven chemistry may be especially important 
for any molecular gas around accreting massive black holes, from
active galaxies in the local Universe (e.g., \citealt{Hailey-Dunsheath2012})
to high-redshift minihalos around the time of cosmic reionization 
\citep{Volonteri2014, Chen2014}.


In this paper we consider 
``one-zone" models for the basic hydrogen, carbon, and oxygen gas-phase formation-destruction chemistry for idealized 
representative gas parcels, and we study the behavior as a function of the overall metallicity, $Z'$, of the gas. We assume that the gas temperatures are fixed by a
radiative heating and cooling balance, but we do not explicitly solve for the 
gas temperature.
As most recently indicated by \citet{Glover2014},
for radiative heating and cooling at low metallicity
a major fraction of the gas mass may rapidly cool
to temperatures of order 100~K via H$_2$ emission line cooling
(see also \citet{Bovino2014} and \citet{Safranek2014}).
In our study we adopt 100~K as our fiducial cold gas temperature.
We wish to understand how the molecular production efficiencies and pathways for the most abundant oxygen and carbon bearing species vary as the metallicity becomes very small,
within the regime of (temperature-insensitive) cold-cloud chemistry.
Thus, the essential parameters are the metallicity $Z'$ (relative to solar) and the
ratio of the driving ionization rate, $\zeta$, to the gas density, $n$.
We also consider the effects of partially attenuated 
background far-ultraviolet (FUV) radiation when or if dust shielding
becomes ineffective, but we do not include any radiative transfer for the FUV.

We presented preliminary results for some of the computations
presented in this paper in \citet{Sternberg2011}. 
There have also been other 
analyses of heavy element molecular chemistry in the early Universe and at low-metallicity (see \citealt{Yan1997, Harwit2003, Dalgarno2006, Vonlanthen2009, Penteado2014}). Discussions of gas phase
chemistry at very high ionization rates have also been presented 
\citep{Lepp1996a, Lepp1998, Bayet2011}
but these have been restricted to local Universe environments,
for solar or near-solar metallicities.
In this paper we consider the low-temperature behavior for a very wide range of metallicities and ionization parameters.

In \S \ref{sec: Network} we present our chemical networks and describe
the dominant formation-destruction reactions and pathways 
We also present a discussion of photoprocesses for partially
shielded gas parcels. In \S \ref{sec: Model ingredients} we write down
and discuss the chemical rate equations, and define our adjustable
parameters, $Z'$, $\zeta/n$, and $I_{\rm UV}/n$. 
In \S \ref{sec: H_H2_analytical} we present an analytic treatment for the atomic to molecular hydrogen balance, and for the steady-state H$_2$ formation time scales. 
In \S \ref{sec: Two illustrative} we develop our analytic scaling relations using 1D numerical model sequences for the chemical behavior as a function of metallicity at fixed ionization parameter, and vice versa. Our focus is on H$_2$, CO, CH, OH, H$_2$O, and O$_2$, and we discuss how the molecular abundance ratios vary with the parameters. 
In \S~\ref{sec: parameter space} we present comprehensive full 2D computations for the chemical behavior, spanning the range from low metallicity to high ionization parameter. 
In \S~\ref{sec:Comparison} we compare our results for ionization-driven 
chemistry in equilibrium gas to the time-dependent behavior 
for low-metallicity collapsing clouds.
We summarize our results and discuss observational implications in \S \ref{sec: Discussion}.

\section{Chemical Networks}
{
\renewcommand\theequation{R\arabic{equation}}
\setcounter{equation}{0}

\label{sec: Network}
We consider simplified (``minimal") interstellar gas phase networks
(Figures \ref{fig: H network}, \ref{fig: OH network}, and \ref{fig: CO network}) for the formation and destruction of H and H$_2$, and for the metal-bearing molecules, OH, H$_2$O, O$_2$, CH, and CO, which are our primary focus. The networks consist of standard ionization-driven two-body ion-molecule sequences, including selective neutral-neutral atom-exchange reactions, and moderated by dissociative and radiative recombination \citep{Herbst1973, Dalgarno1976, vanDishoeck1988, Sternberg1995, vanDishoeck1998, LeTeuff2000a, McElroy2013, Wakelam2012}. We assume that the ionization of H, H$_2$, and He is provided by a flux of penetrating cosmic-rays and/or X-rays, with a total (primary plus secondary) H$_2$ ionization frequency $\zeta$ (s$^{-1}$). We also consider destructive photoprocesses due to background FUV radiation. 

At high metallicities, molecule formation on dust grains
are important for the synthesis of heavy molecules
and their introduction into the gas phase,
especially H$_2$O and O$_2$.
 \citep[e.g.,][]{Cuppen2010, vanDishoeck2013, Lamberts2014}. 
 Furthermore, temperature-dependent freezeout removes molecules from 
 the gas phase, and alters molecular abundance ratios.
 These effects are also sensitive to optical-depth dependent radiation fields,
 as discussed for example by \citet{Hollenbach2009} and \citet{Hollenbach2012}.
However, with decreasing metallicity and dust-to-gas mass ratios, the relative
importance of dust-grain formation of the heavy molecules is expected
to diminish significantly compared to pure gas phase processes.
Furthermore, the intrinsic dust properties are uncertain at
low metallicities, and modeling these requires the introduction
of additional free parameters. In this study we therefore focus on the
well-defined metallicity dependence of just the gas-phase formation-destruction
pathways. Thus, our models are idealized and we caution that for
many realistic environments at high-metallicity our computed
abundances could be modified by the inclusion
of dust interactions.

Negative-ion chemistry \citep{Dalgarno1973, Walsh2009} plays
a role only for H$_2$, via the H$^-$ formation route, as described below. 
Negative ion production reactions for the heavy molecules are included
in our set\footnote{e.g., H$^-$ + O $\rightarrow$ OH + e}
but they remain negligible throughout.
Our focus is on low-temperature $T \lesssim 300$~K gas for which the formation sequences are unaffected by direct H$_2$ neutral-neutral reactions\footnote{e.g., ${\rm O + H_2 \rightarrow OH + H}$; ${\rm OH + H_2 \rightarrow H_2O + H}$} with large endothermicities or activation barriers. Our networks include a total of 74 atomic and molecular hydrogen, carbon, oxygen, nitrogen, silicon, and sulfur species, and a set of 986 reactions, of which  around 80 play a significant role as we describe below.

We wish to study the basic variation trends of the steady-state abundances of the chemical species within isothermal and uniform density ``gas parcels", as functions of the gas phase elemental abundances or metallicity of the gas, the ionization rate, and the total hydrogen gas density. We are particularly interested in the behavior at very subsolar metallicities, and/or high ionization rates. We parameterize the metallicity by a scaling factor $Z'$, such that $Z' = 1$ corresponds to the heavy element abundances in the solar photosphere (\citealt{Asplund2009}) as listed in Table \ref{table: solar abundances}. We consider the solutions to the formation-destruction equations with and without the presence of a moderating and partially attenuated background FUV radiation field. Lyman continuum radiation is always excluded. When a background FUV field is present we assume that for sufficiently high metallicities the gas parcels are fully shielded by surrounding optically thick columns of dust. As $Z'$ becomes small we allow the gas to become optically thin to dust-absorption. However, we assume that the parcels are always shielded by sufficiently large columns of H$_2$ such that the 912-1108~\AA $\,$ Lyman-Werner (LW) band is fully blocked by fully overlapping H$_2$ absorption lines. We introduce this assumption to allow for optimal conditions for molecule formation in the parcels even at very low metallicity.

\begin{table}
\caption{Heavy element abundances relative to hydrogen.} 
\centering 
\begin{tabular}{c c}
\\ [-1.5ex]
\hline\hline 
Element & \; $n_i/n$ \\ [0.5ex] 
  \hline 
C & $2.9 \times 10^{-4}$ \\
N & $6.8 \times 10^{-4}$ \\
O & $4.9 \times 10^{-4}$ \\
Si & $3.2 \times 10^{-5}$ \\
S & $1.3 \times 10^{-5}$ \\
\hline 
\end{tabular}
\label{table: solar abundances} 
\end{table}

In our description of the networks we describe the dominant formation and 
destruction pathways in the limits of high and low metallicity. As we discuss in 
\S\ref{sec: H_H2_analytical} and \S\ref{sec: Two illustrative}, for any $Z'$ the hydrogen makes the transition from predominantly atomic to molecular form at a critical ratio of the ionization rate $\zeta$ to the total hydrogen gas density $n$.
In our discussion, the high-metallicity formation-destruction pathways usually (but not always) correspond to predominantly H$_2$ gas. The low-metallicity pathways usually correspond to predominantly H gas. The transition from the fully H$_2$ ``molecular regime" to the fully H ``atomic regime" is important for the chemical behavior, as we describe below.

We begin with a description of the hydrogen/helium networks, and then consider the oxygen, and
carbon/oxygen sequences.

\subsection{Hydrogen and Helium}
\label{sub: H and He}
We illustrate the hydrogen-helium networks in Figure \ref{fig: H network}.
These networks include the species H, H$_2$, H$^+$, H$^-$, H$_2^+$, H$_3^+$, He, He$^+$, and HeH$^+$.

The formation of H$_2$ is essential for the efficient ion-molecule production of metal-bearing molecular species in the gas phase. At high metallicity H$_2$ formation is dominated by dust-grain catalysis
\begin{equation}
\label{R: H2 form dust}
{\rm  H \; + \; H:gr   \; \rightarrow  \;  \Ht  \;  +  \;  gr}	\; \; \; .	
\end{equation}
At sufficiently low metallicity the main H$_2$ formation route is in the gas phase via the formation of H$^-$ negative ions
\begin{equation}
\label{R: H- form}
{\rm H  \; + \;  e  \; \rightarrow \;  H^- \;  + \;  \nu}	
\end{equation}
\begin{equation}
\label{R: H2 form gas}
{\rm H^- \;  + \;  H \;  \rightarrow \;  H_2  \; + \; e} \; \; \; .	
\end{equation}
This radiative-attachment associative-detachment sequence is moderated by mutual neutralization
\begin{equation}
\label{R: H- dist H+}
{\rm H^-  \; + \;  H^+ \;  \rightarrow  \; H \; + \; H}	 \; \; \; ,
\end{equation}  which limits the H$^-$ abundances when the proton density becomes large.
Gas phase H$_2$ production also proceeds via radiative association
\begin{equation}
\label{R: H2+ form by H+}
{\rm H \; + \; H^+ \; \rightarrow \; H_2^+ \; + \; \nu \; \; \; ,}
\end{equation}
followed by charge transfer
\begin{equation}
\label{R: H2 form by H2+}
{\rm H_2^+ \; + \; H \; \rightarrow \; H_2 \; + \; H^+ \; \; \; .}
\end{equation}
However, at the low gas temperatures ($T \lesssim 300$K) we are considering, 
and in the absence of photodetachment, the H$^-$ sequence, [\ref{R: H- form}] and [\ref{R: H2 form gas}], always dominates.

In the absence of FUV photodissociation, the H$_2$ is removed by (primary plus secondary)
cosmic-ray and/or X-ray ionization and dissociative ionization. We refer to these processes collectively as ``crx-ionization". Thus
\begin{align}
\label{R: H2 dist cr form H2+}
 {\rm H_2 \;  +  \; crx \;  }&\rightarrow {\rm \;  H_2^+  \; + \; e} \\
 \label{R: H2 dist cr form H+}
 &\; {\rm \rightarrow \;  H^+ \; +  \; H  \; + \;  e \; \; \; .}
\end{align}
Depending on the H/H$_2$ density ratio, the H$_2^+$ ions are removed by charge transfer (reaction [\ref{R: H2 form by H2+}]), leading back to H$_2$, and by proton transfer
\begin{equation}
\label{R: H3+ form}
{\rm H_2^+ \;  +  \; H_2 \;  \rightarrow   \; H_3^+ \;  + \;  H}                                                                                           \; \; \; ,
\end{equation} leading to H$_3^+$.
The H$_3^+$ ions are removed by dissociative recombination
\begin{align}
\label{R: H3+ recombiantion 3H}
{\rm H_3^+ \; + \; e  \;} &{\rm \rightarrow \;  H \; + \; H \; + \; H} \\
\label{R: H3+ recombination H2 H}
&{\rm \rightarrow \;  H_2 \; + \; H \; \; \; ,}
\end{align} and also by
\begin{equation}
\label{R: H3+ dist with CO}
{\rm H_3^+ \; + \; CO \; \rightarrow \;  HCO^+ \; + \; H}
\end{equation} when the e/CO ratio is small.

The H/H$_2$ density ratio is a critical parameter affecting the overall chemical behavior. The main sources of free hydrogen atoms are direct
dissociative ionization of the H$_2$, or H$_2$ ionization ([\ref{R: H2 dist cr form H+}] and [\ref{R: H2 dist cr form H2+}]) followed by proton-transfer formation of H$_3^+$ ([\ref{R: H3+ form}]).
At high metallicity, the H atoms are removed by the H$_2$ grain surface formation process. At low metallicity the H atoms are removed by ionization
\begin{equation}
\label{R: H dist form H+}
{\rm H  \; + \;  crx \;  \rightarrow \;  H^+ \;  + e} \; \; \; .	
\end{equation}
The ionization of H is a major source of protons at low metallicity where the atomic fraction becomes large. When the gas is primarily H$_2$,
the protons are produced by dissociative ionization of the H$_2$ (reaction [\ref{R: H2 dist cr form H+}]). At high metallicity the protons are removed mainly by charge transfer with atomic oxygen
\begin{equation}
\label{R: O+ form}
{\rm H^+ \; + \; O  \; \rightarrow \;  H \; + \; O^+ \; \; \; .}
\end{equation} Reactions with molecules
\begin{align}
\label{R: H+ dist OH}
{\rm H^+ \; + \; OH  \;} &{\rm \rightarrow \;  OH^+ \; + \;  H}\\
\label{R: H+ dist O2}
{\rm H^+ \; + \; SiO  \;} &{\rm \rightarrow \; H \; + \;  SiO^+ \; \; \; ,}
\end{align} may also contribute to the removal of H$^+$ in some parts of our parameter space.
At low metallicity, radiative recombination
\begin{equation}
\label{R: recombination}
{\rm H^+ \; + \; e  \; \rightarrow \;  H \; + \; \nu \; \; \; ,}
\end{equation} is the dominant H$^+$ removal mechanism.

As we discuss further below, ionization of atomic hydrogen is the major source of free electrons at low metallicity,
and the protons then carry all of the positive charge.
At high metallicity, electrons are provided by ionization of heavy elements, and the
positive charge is carried by metal ions and/or molecular ions.
 \begin{figure}
 \includegraphics[width=.43\textwidth]{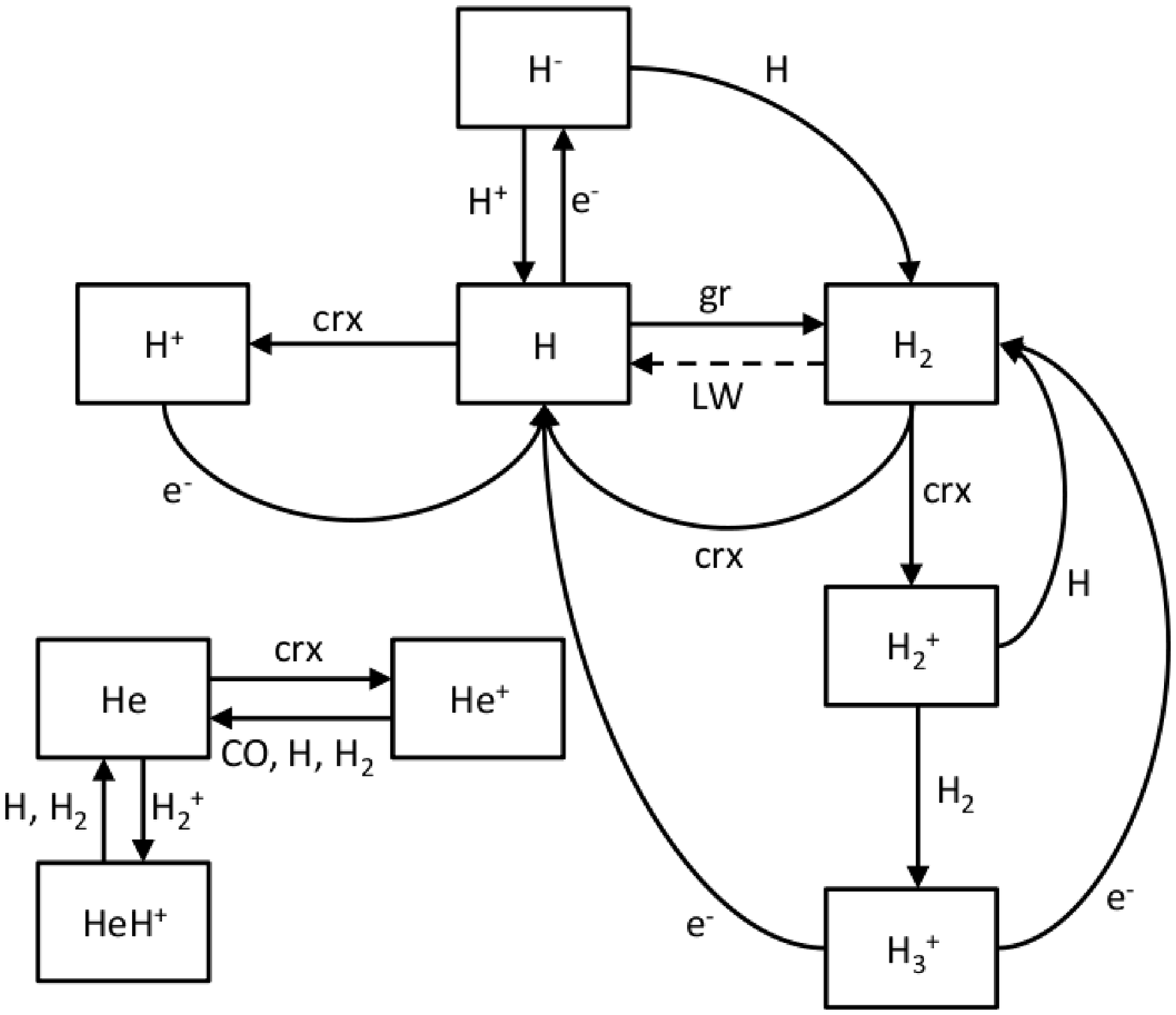}
\caption{Hydrogen and helium reaction networks.}
\label{fig: H network}
 \end{figure}

Helium atoms and ions interact with hydrogen species and also with CO molecules. He$^+$ is produced by crx-ionization
\begin{equation}
\label{R: He+ form}
{\rm He \;  + \;  crx \;  \rightarrow \;  He^+ \;  + \;  \nu}                                                                                                                                                                             \; \; \; .
\end{equation} At high metallicity, He$^+$ is neutralized by dissociative charge transfer with CO
\begin{equation}
\label{R: He+ dist High Z}
{\rm He^+ \;  + \;  CO \;  \rightarrow  \; He \;  + \;  O \;  +  \; C^+ \; \; \; .}
\end{equation}
At low metallicity, the He$^+$ ions are removed rapidly by charge transfer and dissociative charge transfer with H and H$_2$
\begin{align}
\label{R: He+ dist H}
{\rm He^+ \;  + \;  H  \; } &{\rm \rightarrow \;  He \;  + \;  H^+} \\
\label{R: He dist H2 to He H2+}
{\rm He^+ \; + \; H_2 \; }  &{\rm \rightarrow  \; He \; + \; H_2^+}\\
\label{R: He dist H2 to He H+ H}
& {\rm \rightarrow  \; He \; + \; H^+ \;  + H  \;  \; \; .}
\end{align} 
Because these reactions are rapid, the He$^+$ fraction remains small.
Finally, the molecular ion HeH$^+$ is produced via
\begin{equation}
{\rm He  \; +  \; H_2^+  \; \rightarrow  \; HeH^+ \;+ \;  H \; \; \; ,}
\end{equation} and is removed by
\begin{equation}
{\rm HeH^+  \; +  \; H_2 \;  \rightarrow  \; He \;  +  \; H_3^+} \; \; \; ,
\label{R: HeH+ diss highZ}
\end{equation} or by
\begin{equation}
{\rm HeH^+ \;  + \;  H  \; \rightarrow \;  He \;  + \;  H_2^+ }                                                                                                                                                                              \; \; \; ,
\label{R: HeH+ diss lowZ}
\end{equation} depending on the H/H$_2$ ratio.
In our parameter space
the formation of HeH$^+$ via radiative association
\begin{equation}
{\rm He \; + \; H^+ \; \rightarrow \; HeH^+ \; + \; \nu} \; \; \; ,
\end{equation}
followed by reactions [R25] and [R6] is a negligible source of H$_2$
compared to the H$^-$ formation route. But in the recombination era,
prior to the appearance of ionization sources and with the rapid removal
of the H$^-$ by the thermal background radiation (at redshifts $z \gtrsim 100$), this was a major
source of H$_2$, as was the sequence [R5]-[R6].

\subsection{Oxygen: OH, H$_2$O, and O$_2$}
\label{sub: OH H2O}
Our oxygen network includes O, O$^+$, OH$^+$, H$_2$O$^+$, H$_3$O$^+$, OH, H$_2$O and O$_2$, and also silicon-oxygen and sulfur-oxygen species, and the carbon-oxygen branches described in \S\ref{sub: C}. 
In Figure \ref{fig: OH network} we show the oxygen-hydride portion of the network, with our focus on the formation and destruction of OH, H$_2$O, and O$_2$.
 At temperatures $\lesssim 300$~K, the production of OH and H$_2$O is initiated by charge transfer
 \begin{align}
\tag{\ref{R: O+ form}}
{\rm O \; + \; H^+  \;} &{\rm \leftrightarrows  \; O^+ \; + \; H}\\
{\rm O^+ \; + \; H_2  \;} &{\rm \rightarrow \;  OH^+ \; + \; H}
\label{R: OH+ form}
\end{align} or by proton transfer
\begin{equation}
\label{R: OH+ form with H3+}
{\rm O  \; + \;  H_3^+  \; \rightarrow OH^+ \; + \; H_2}                                                                           \; \; \; ,
\end{equation} followed by the rapid abstractions
\begin{align}
\label{R: H2O+ form}	
{\rm OH^+ \;  + \; H_2 \;} &{\rm \rightarrow \;  H_2O^+ \;  + \; H} \\
\label{R: H3O+ form} \; \; \; 
{\rm H_2O^+  \; + \; H_2 \;} &{\rm \rightarrow \;  H_3O^+  \; + \; H \; \; \; .}
\end{align} The sequence is terminated by dissociative recombination
 \begin{figure}
  \includegraphics[width=0.35\textwidth]{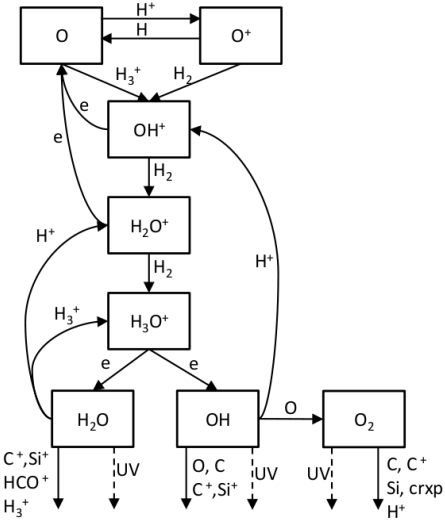}
\caption{Ion-molecule formation-destruction pathways for OH, H$_2$O and O$_2$.}
 \label{fig: OH network}
 \end{figure}
\begin{align}
\label{R: OH form 1 (OH,H,H)}
{\rm H_3O^+ \; + \; e \;}  &\rightarrow  {\rm \; OH \; + \; H \; + \; H}  \\
\label{R: OH form 2 (OH,H2)}
{\rm H_3O^+ \; + \; e \;}  &\rightarrow  {\rm \; OH \; + \; H_2}  \\
\label{R: H2O form}
& \; {\rm\rightarrow  \; H_2O \;  + \;  H } \\
\label{R: H3O+,e to O}
& \; {\rm\rightarrow  \; O \;  + \; H_2 \; + \; H \; \; \; , }
\end{align}
which yields the OH and H$_2$O molecules. When the fractional ionization is large, the abstraction sequence 
may be interrupted by dissociative recombination of the reactive ions
\begin{align}
\label{R: H2O+ dist}
{\rm H_2O^+ \;  +  \; e  \;} &\rightarrow {\rm \;  OH \; + \; H  } \\
\label{R: H2O+ dist 2}
&\; {\rm            \rightarrow O \; + \; H_2  } \\
\label{R: H2O+ dist 3}
&\; {\rm            \rightarrow O \; + \; H \; + \; H  }
\end{align} and
\begin{equation}
{\rm OH^+ \;  +  \; e \;  \rightarrow \;  O \; + \; H  }                                                                             \; \; \; .
\label{R: OH+ dist}
\end{equation}
Importantly, at high metallicities OH is removed mainly by rapid reactions with oxygen atoms
\begin{align}
\label{R: OH dist high Z with O}
{\rm OH \;  + \;  O \;} &{\rm  \rightarrow \;  O_2 \;   +  \; H  } \; \; \; .
\end{align}
Additional removal reactions are
\begin{align}
\label{R: OH dist high Z with C}
{\rm OH  \; + \;  C  \;} &{\rm \rightarrow  \; CO \;   +  \; H  \; \; \;}\\
\label{R: OH dist high Z with C+}
{\rm OH  \; + \;  C^+ \; }&{\rm \rightarrow \;  CO^+ \;   + \;  H  } \\ 
\label{R: OH dist high Z with Si+} 
{\rm OH  \; + \;  Si^+ \;} &{\rm \rightarrow \;  SiO^+  \;  + \;  H  \; \; \; .}
\end{align}
At low metallicity, OH is removed mainly by charge transfer with protons
\begin{equation}
{\rm OH \;  + \;  H^+ \;  \rightarrow  \; OH^+  \;  +  \; H  }                                                                            \; \; \; . \label{R: OH dist lowZ}
\end{equation}
Reaction [\ref{R: OH dist lowZ}] is a destruction channel because some of the OH$^+$, H$_2$O$^+$, and H$_3$O$^+$ ions undergo dissociative-recombination leading directly to atomic oxygen and to the complete breakup of the chemical bonds. The transition from OH removal by oxygen atoms to removal by protons is crucial for the behavior of the OH abundances with varying metallicity and ionization parameter, as we describe in detail in \S\ref{sec: Two illustrative} and \S\ref{sec: parameter space}.

In addition to OH, the dissociative recombination of H$_3$O$^+$ yields H$_2$O molecules (reaction [\ref{R: H2O form}]). At high metallicities the H$_2$O is removed by rapid reactions with metal species
\begin{align}
\label{R: H2O dist with C+}
{\rm H_2O  \; +  \; C^+  \;} &{\rm \rightarrow  \; HCO^+ \; + \;  H  }\\
 \label{R: H2O dist with Si+}
{\rm H_2O \;  +  \; Si^+ \;} &{\rm \rightarrow \;  SiOH^+ \;   +  \; H }\\
\label{R: H2O dist with HCO+}
{\rm H_2O \;  +  \; HCO^+ \;} &{\rm \rightarrow \;  H_3O^+ \; + \;  CO  \; \; \; ,}
\end{align} but it is also removed by
\begin{equation}
{\rm H_2O  \; + \;  H_3^+ \;  \rightarrow \;  H_3O^+  \;  + \;  H_2  \; \; \; .}
\label{R: H2O dist with H3+}
\end{equation}
At low metallicity the H$_2$O is removed by
\begin{equation}
{\rm H_2O  \; +  \; H^+  \; \rightarrow \;  H_2O^+  \;  + \;  H  }
\label{R: H2O dist with H+} \; \; \; .
\end{equation}
This is a removal mechanism for the H$_2$O (as is [\ref{R: H+ dist OH}] for OH) because some of the H$_2$O$^+$ ions are removed by dissociative recombination rather than reentering the abstraction sequence that leads back to H$_2$O. 

Following the formation of OH via [\ref{R: O+ form}], [\ref{R: OH+ form}]-[\ref{R: OH form 2 (OH,H2)}],
O$_2$ is produced by
\begin{equation}
{\rm OH \; + \; O \; \rightarrow \; O_2 \; + \; H \; \; \; ,}
\tag{\ref{R: OH dist high Z with O}}
\end{equation}
which as stated above is also the primary OH removal reaction at
high metallicity.
At high metallicity, the O$_2$ is removed by
\begin{align}
\label{R: O2 dist with C (form CO)}
{\rm O_2  \; +  \; C  \;} &{\rm \rightarrow \;  CO  \;  + \;  O  }\\
 \label{R: O2 dist with C+ (form CO)}
{\rm O_2  \; +  \; C^+  \;} &{\rm \rightarrow \;  CO  \;  + \;  O^+  }\\
 \label{R: O2 dist with C+ (form CO+)}
&{\rm \rightarrow \;  CO^+  \;  + \;  O  }\\
 \label{R: O2 dist with Si (form SiO)}
{\rm O_2  \; +  \; Si  \;} &{\rm \rightarrow \;  SiO  \;  + \;  O  } \; \; \; .
\end{align} The O$_2$ is also removed by internal photodissociation
induced by secondary electron excitations of the H$_2$
 (\citealt{Sternberg1987}; \citealt{Gredel1989}; \citealt{Heays2014}).
 Here and below, we refer to induced photodissociation with the label ``crxp". Thus,
\begin{align}
 \label{R: O2 dist with crp}
{\rm O_2  \; +  \; crxp  \;} &{\rm \rightarrow \;  O  \;  + \;  O  }
 \; \; \; .
\end{align}
At low metallicity, the O$_2$ is removed by protons in the sequence
\begin{align}
\label{R: O2 dist with H+ (form O2+)}
{\rm O_2  \; +  \; H^+  \;} &{\rm \rightarrow \;  O_2^+ \; + \; H }\\
 \label{R: O2+ dist with e}
{\rm O_2^+  \; +  \; e \;} &{\rm \rightarrow \;  O  \;  + \;  O  } \; \; \; ,
\end{align}
rather than by reactions with metal atoms and ions.

When any penetrating FUV radiation is present direct photodissociation and photoionization can become important in reducing the OH, H$_2$O, and O$_2$ abundances. We discuss the FUV photoprocesses further in \S\ref{sub: Photoreactions}.

\subsection{Carbon: CH and CO}
\label{sub: C}

 \begin{figure}
 \includegraphics[width=0.43\textwidth]{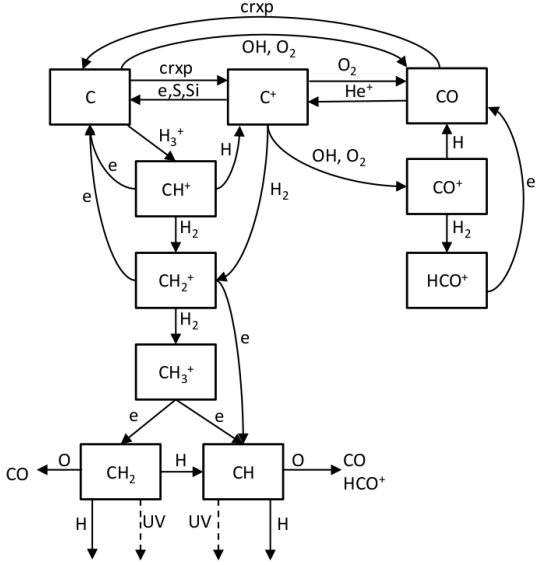}
\caption{Formation-destruction pathways for CH and CO.}
   \label{fig: CO network}
 \end{figure}
 
Our carbon-oxygen network is shown in Figure \ref{fig: CO network}. It includes C, C$^+$, CH$^+$, CH$_2^+$, CH$_3^+$, CH, CH$_2$, CO$^+$, HCO$^+$, and CO. We are mainly interested in the formation and destruction of CH and CO.

The production of CH is initiated by the formation of CH$_2^+$ via proton transfer and abstraction
\begin{align}
\label{R: CH+ form}
{\rm C  \; +  \; H_3^+  \; \rightarrow} &{\rm \;  CH^+ \;  + \;  H_2  }\\
\label{R: CH2+ form from CH+ H2}
{\rm CH^+  \; + \;  H_2 \;  \rightarrow} &{\rm \; CH_2^+  \; + \;  H } \; \; \; , 	
\end{align} or by direct radiative association
\begin{equation}
{\rm C^+  \; + \;  H_2 \;  \rightarrow  \; CH_2^+ \;  + \;  \nu    } 	 \; \; \; .
	\label{R: CH2+ form from C+ H2}
\end{equation} The relative efficiencies of [\ref{R: CH+ form}]-[\ref{R: CH2+ form from CH+ H2}] versus [\ref{R: CH2+ form from C+ H2}] depend on the C$^+$/C ratio. The formation of CH$_2^+$ is followed by
\begin{equation}
{\rm CH_2^+  \; + \;  H_2 \;  \rightarrow \;  CH_3^+ \;  +  \; H    } \; \; \; ,
\label{R: CH3+ form}
\end{equation} and then dissociative-recombination
\begin{align}
\label{R: CH2 form}
{\rm CH_3^+ \;  + \;  e \;} &\rightarrow {\rm \; CH_2 \;  + \;  H    } \\
\label{R: CH form from CH3+}
&\; {\rm \rightarrow  \; CH \; + \; H_2}	\\	\label{R: CH form from CH3+ 2}
&\; {\rm \rightarrow  \; CH \;  +  \; H \;  +  \; H    }                                                                         \; \; \; ,
\end{align} terminating in the production of CH (and CH$_2$). 
When the fractional ionization is large the abstraction sequence is moderated by
\begin{align}
\label{R: CH+ dist with e}
{\rm CH^+  \; + \;  e \; } &{\rm \rightarrow \;  C \;  + \;  H \; \; \;    } \\
\label{R: CH2+ dist with e form CH}
{\rm CH_2^+  \; + \;  e \;}  &\rightarrow {\rm \;  CH \;  + \;  H   \; \; \; } \\ 	
\label{R: CH2+ dist with e 2}
&\; {\rm \rightarrow \;  C \;  + \;  H  \; + \; H \; \; \; } \\
\label{R: CH2+ dist with e 3}
&\; {\rm \rightarrow \;  C \;  + \;  H_2   \; \; \; , } 	
\end{align} and
\begin{equation}
{\rm CH^+ \;  + \;  H \;  \rightarrow \;  C^+ \;  + \;  H_2    } \; \; \; .	\label{R: CH+ dist with H}
\end{equation} Reaction [\ref{R: CH2+ dist with e form CH}] is an additional source of CH, but overall these moderating reactions tend to reduce the CH formation efficiency. When the e/H ratio is small
\begin{equation}
{\rm CH_2  \; + \;  H \;  \rightarrow \;  CH \;  +  \; H_2    } \; \; \; ,
\label{R: CH form from CH2}
\end{equation} also contributes to CH formation.

In most of our parameter space, the dominant CH destruction reaction is
\begin{equation}
{\rm CH  \; +  \; H  \;  \rightarrow  \; C  \; +  \; H_2   } 	 \; \; \; .	\label{R: CH dist}
\end{equation} This, in contrast with OH, which is not removed by H atoms\footnote{The binding energies of H$_2$, OH, and CH are 4.48, 4.41 and 3.49 eV respectively.
The reaction OH + H $\rightarrow$ O + H$_2$ is exothermic, but it has a large barrier
\citep{Balakrishnan2004, vanDishoeck2013} and is ineffective in cold gas.
CH + H $\rightarrow$ C + H$_2$ is much more energetically favorable, and is rapid at
low temperatures \citep{Grebe1982, Harding1993}.}. The destruction efficiency increases with the atomic hydrogen fraction, and CH therefore disappears as the metallicity is reduced. We discuss this further in 
\S\ref{sec: Two illustrative} where we show that CH vanishes compared to OH
as the metallicity is reduced.

We now consider CO, which is produced via several channels.
First is via OH,
\begin{equation}
{\rm C  \; +  \; OH \;  \rightarrow \;  CO \;  + \;  H \; \; \; ,}
\label{R: CO form from C,OH}
\end{equation} or
\begin{align}
\label{R: CO+ form}
{\rm C^+ \;  + \;  OH \;} &{\rm  \rightarrow  \; CO^+ \;  + \;  H    }\\
\label{R: CO form from CO+ H low Z}
{\rm CO^+ \;  + \;  H \;} &{\rm \rightarrow  \; CO  \; +  \; H^+    }  \; \; \; ,
\end{align}
and
\begin{align}
\label{R: HCO+ form}
{\rm CO^+ \;  + \;  H_2 \;} &{\rm  \rightarrow \;  HCO^+ \;  + \;  H    }\\
\label{R: CO form from HCO+ high Z}
{\rm HCO^+  \; + \;  e  \;} &{\rm \rightarrow  \; CO  \; + \;  H  \; \; \; .  }
\end{align} Reactions [\ref{R: CO form from C,OH}]-[\ref{R: CO form from HCO+ high Z}] are the CO production pathways via the ``OH-intermediary". 
For a given OH abundance, the relative efficiencies of these reactions depend on the C$^+$/C and H/H$_2$ density ratios. 
As we will discuss below, CO production via OH always dominates in the low $Z'$ limit.

At sufficiently high $Z'$, the CH/OH ratio may become large (see \S\ref{sub: Z dependence}), and CO is then also produced by 
\begin{align}
\label{R: CO form from CH}
 {\rm O \;  + \;  CH \;} & {\rm \rightarrow  \; CO  \; +  \; H} \\                                                                                                        
\label{R: CO form from CH2 (H2)}
{\rm O \;  + \;  CH_2 \;} &{\rm \rightarrow  \; CO  \; +  \; H_2 } \\                                                                                  
\label{R: CO form from CH2 (H H)}
&{\rm \rightarrow  \; CO  \; +  \; H \; + \; H }                                                                                  
\; \; \; .
\end{align}
Another route is via chemionization
\begin{align}
\label{R: HCO+ form from CH (O)}
{\rm O \;  + \;  CH \;} &{\rm \rightarrow \; HCO^+  \; +  \; e}\\
\tag{\ref{R: CO form from HCO+ high Z}}
{\rm HCO^+ \;  + \;  e \; }&{\rm \rightarrow  \; CO \;  + \; H \; \; \;. } 
\end{align} 
Reactions [\ref{R: CO form from HCO+ high Z}]-[\ref{R: HCO+ form from CH (O)}] are the CO production pathways via the ``CH intermediary". 
At sufficiently high metallicity (but mostly outside our parameter space) these reactions also contribute to the removal of the CH in addition to removal by H atoms.

CO formation can also proceed via
\begin{align}
\tag{\ref{R: O2 dist with C (form CO)}}
{\rm O_2  \; +  \; C  \;} &{\rm \rightarrow \;  CO  \;  + \;  O  }\\
 \tag{\ref{R: O2 dist with C+ (form CO)}}
{\rm O_2  \; +  \; C^+  \;} &{\rm \rightarrow \;  CO  \;  + \;  O^+  }\\
 \tag{\ref{R: O2 dist with C+ (form CO+)}}
&{\rm \rightarrow \;  CO^+  \;  + \;  O  } \; \; \;,
\end{align} 
followed by [\ref{R: CO form from CO+ H low Z}]-[\ref{R: CO form from HCO+ high Z}]. These reactions are the CO production pathways via the ``O$_2$ intermediary". 
As we discuss below,
CO formation via O$_2$ becomes important at low $\zeta/n$ in the molecular regime where the O$_2$/OH abundance ratio can become large.


In the absence of FUV photodissociation, CO is destroyed mainly by dissociative charge transfer with He$^+$,
\begin{equation}
{\rm He^+ \;  + \;  CO \;  \rightarrow  \; He \;  + \;  O \;  +  \; C^+ \tag{\ref{R: He+ dist High Z}}} \; \; \;  .
\end{equation} This is the primary CO removal mechanism at all metallicities.

Reaction [\ref{R: He+ dist High Z}] is a major source of C$^+$ when much of the carbon is locked in CO, as occurs at high-metallicity. An additional source of C$^+$, at all metallicities, is via crxp-ionization of any free carbon atoms
\begin{equation}
{\rm C  \; + \;  crxp \;  \rightarrow \;  C^+ \;  +  \; e       }                                                                                                                     \; \; \; .
\label{R: C+ form by crp}
\end{equation} 
The C$^+$ ions are removed by radiative recombination
\begin{equation}
{\rm C^+  \; + \;  e \;  \rightarrow  \; C \;  + \;  \nu    }                                                                 \; \; \; ,
\label{R: C+ recombination}
\end{equation} charge transfer
\begin{align}
\label{R: C+ dist with metals 1}
{\rm C^+ \;  +  \; S  \; } &{\rm \rightarrow \;  C \;  + \;  S^+ }\\
\label{R: C+ dist with metals 2}
{\rm C^+  \; + \;  Si \; } &{\rm \rightarrow  \;  C  \; + \;  Si^+    \; \; \; ,}
\end{align} radiative association (reaction [\ref{R: CH2+ form from C+ H2}]), and via the CO-forming reactions with OH or O$_2$ (reaction [\ref{R: CO+ form}] and [\ref{R: O2 dist with C+ (form CO)}]-[\ref{R: O2 dist with C+ (form CO+)}]).

At high metallicity reactions [\ref{R: C+ dist with metals 1}], [\ref{R: C+ dist with metals 2}], and induced photodissociation
\begin{equation}
{\rm CO  \; + \;  crxp \;  \rightarrow \;  C \;  +  \; O       }                                                                                                                     \; \; \; ,
\label{R: C form by CO crp}
\end{equation} dominate the production of free atomic carbon. 
The carbon atoms are removed by induced ionization (reaction [\ref{R: C+ form by crp}]), by
\begin{equation}
{\rm C  \; + \;  H_3^+ \;  \rightarrow \;  CH^+ \;  +  \; H_2       }                                                                                                                     \; \; \; ,
\tag{\ref{R: CH+ form}}
\end{equation} and also by the CO-forming reactions with OH and O$_2$
(reactions [\ref{R: CO form from C,OH}] and [\ref{R: O2 dist with C (form CO)}]).

At low metallicity the C$^+$/C ratio is set by the balance between induced photoionization [\ref{R: C+ form by crp}] and radiative-recombination [\ref{R: C+ recombination}].
\subsection{Photoprocesses}
\label{sub: Photoreactions}
We also present computations including the effects of photodissociation and photoionization by externally incident FUV radiation. At very low metallicities, dust shielding may be ineffective, and photoprocesses will become significant in the presence of FUV. However, we assume that even at low $Z'$, the gas parcels are always shielded by H$_2$ gas columns $\gtrsim 10^{22}$~cm$^{-2}$ that completely block the 918 -1108 \AA$\,$ LW band via fully overlapping H$_2$ absorption lines (e.g.~\citealt{Sternberg2014}).

  \begin{figure}
 \includegraphics[width=.5\textwidth]{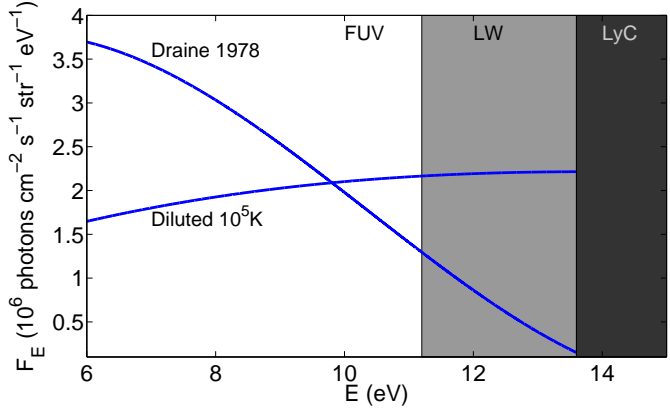}
\caption{The Draine FUV spectrum and our diluted 10$^5$~K black-body FUV spectrum, both for $I_{\rm UV}=1$ (see text). The light shaded region is the 11.2-13.6 eV LW band.}

 \label{fig: radiation spectrum}
 \end{figure}

\ctable[
star,
cap = Photorates,
caption = Photorates for \mbox{$I_{\rm UV}=1$},
label = table: photorates,
]{l c c c c c}{
\tnote{CO photodissociation occurs via absorption-line predissociation \citep{Visser2009} and 11.5 eV is the lowest photon energy in this multiline process.}
\tnote[b]{In computing the photodeatachment rate we adopt the normalized Draine and diluted 10$^5$~K photon intensities from 13.6 eV all the way to the H$^-$ electron detachment threshold of 0.75 eV.}
}{ \FL
Reaction & Threshold (eV) & \multicolumn{4}{c}{ Photorate \mbox{$\Gamma \, (10^{-10} \, {\rm s}^{-1})$}} \ML

 $\;$ & $\;$ & \multicolumn{2}{c}{Draine} & \multicolumn{2}{c}{Diluted 10$^5$~K}   \NN
 & & thin &  LW-blocked  & thin &  LW-blocked \ML
${\rm OH \; + \;  \nu \;  \rightarrow \;  O \;  + \;  H    }                                                                 $ & 6.4 & 3.8  & 2.8 & 4.7 & 2.5 \NN
${\rm  H_2O  \; + \;  \nu \;  \rightarrow  \; O \;  + \;  H_2    }                                                                $ & 9.5 & 0.49 &  0.28 & 1.1 & 0.32\NN
${\rm  H_2O  \; + \;  \nu \;  \rightarrow  \; OH \;  + \;  H    }                                                                $ & 6.0 & 7.5 &  5.5 & 11.7 & 4.8 \NN
${\rm O_2  \; + \;  \nu  \; \rightarrow \;  O  \; +  \; O    }                                                                $ & 7.0 & 7.9 & 7.0 & 9.4 & 5.1 \NN
${\rm CH \;  + \;  \nu  \; \rightarrow \;  C \;  +  \; H    }                                                                  $ & 3.4 & 9.0 & 8.8 & 5.5 & 4.7 \NN
${\rm CO \;  + \;  \nu  \; \rightarrow \;  C \;  +  \; O    }                                                                  $ & 11.5\tmark\ & 2.6 & 0.0 & 14.2 & 0.0 \NN
${\rm C  \; + \;  \nu  \; \rightarrow \;  C^+ \;  + \;  e    }                                                                 $ & 11.3 & 3.2 & 0.0 & 10.1 & 0.0 \NN
${\rm CH  \; + \;  \nu  \; \rightarrow \;  CH^+ \;  + \;  e    }                                                                 $ & 3.4 & 7.7 & 0.97 & 21.3 & 1.4 \NN
${\rm H^- \;  + \;  \nu \;  \rightarrow \;  H \;  + \;  e    }                                                                 $ & 0.75\tmark[b]\ & 55.8 & 54.2 & 27.7 & 22.9 \LL

}

For such conditions, species such as C or CO with photodestruction thresholds within the LW band are fully shielded against the FUV.
But for species with thresholds longward of 1108~\AA , removal by FUV radiation can become important. This includes photodissociation
\begin{align}
\label{R: FUV OH dist}
{\rm OH \; + \;  \nu \;} &{\rm \rightarrow \;  O \;  + \;  H}\\
\label{R: FUV H2O dist}
{\rm H_2O  \; + \;  \nu \;}  &{\rm \rightarrow  \;  OH \;  + \;  H} \\
\label{R: FUV H2O dist 2}
&{\rm \rightarrow  \; O \;  + \;  H_2}\\
\label{R: FUV O2 dist}
{\rm O_2  \; + \;  \nu  \;} &{\rm \rightarrow \;  O  \; +  \; O}\\
\label{R: FUV CH dist}
{\rm CH \;  + \;  \nu  \;} &{\rm \rightarrow \;  C \;  +  \; H \; \; \; ,}                                                                
\end{align} photoionization
\begin{align}
\label{R: FUV CH dist 2}
{\rm CH  \; + \;  \nu  \;} &{\rm \rightarrow \;  CH^+ \;  + \;  e \; \; \; ,   }                                                                 
\end{align} and photodetachment
\begin{equation}
{\rm H^- \;  + \;  \nu \;  \rightarrow \;  H \;  + \;  e    }                                                                 \; \; \; .
\label{R: FUV H- dist}
\end{equation}
We have used the photoprocess cross-sections in the 
\citet{vanDishoeck2006} database\footnote{http://home.strw.leidenuniv.nl/$\sim$ewine/photo/} to calculate the dust-free photodissociation and photoionization rates for our species set assuming the \citet{Draine1978, Draine2011} representation for the Galactic interstellar radiation field, as well as for a diluted 10$^5$ K blackbody spectrum. 
For H$^-$ we used the photodetachment cross section calculated by
\citet[][and H.~Sadeghpour and P.~Stancil, private communication]{Miyake2010}. 
Computations of the H$^-$ photodetachment rate
\footnote{ Latif et al.~presented H$^-$ photodetachment rates for $J_{21}=1$, where $J_{21}$ is the specific intensity in units of 10$^{-21}$ erg s$^{-1}$ cm$^{-2}$ sr$^{-1}$ Hz$^{-1}$, at the Lyman limit. For a $T=10^5$~K blackbody spectrum $I_{\rm UV} = 182 J_{21}$, and our detachment rate of $2.7\times10^{-9} I_{\rm UV}$~s$^{-1}$ is consistent with their result (see their Figure 1).}
 as a function of radiation temperature for blackbody spectra have also been presented by \citet{Latif2015}.

We adopt the 10$^5$~K spectrum as representative of 
the background FUV produced by massive Pop-III stars. We normalize the free-space fields by a scaling factor $I_{\rm UV}$ such that for $I_{\rm UV}=1$, the total photon density in the 6-13.6 eV band is $6.5 \times 10^{-3}$~cm$^{-3}$, as for the unit free space Draine field. The spectral shapes of the two 
normalized radiation fields are displayed in Figure \ref{fig: radiation spectrum}. In Table \ref{table: photorates} we assemble a selection of our computed list of photorates for $I_{\rm UV}=1$, for the Draine and the diluted blackbody fields assuming the LW-band is either optically thin or fully blocked (and with Lyman-continuum radiation always excluded). 
The photorates listed in Table \ref{table: photorates} are the important $b_{ij}$ factors that enter into our Equations (\ref{eq: chemical equilibrium IUV on}) and (\ref{eq: chemical equilibrium x IUV on}) below.

}

\section{Rate Equations}
\renewcommand\theequation{\arabic{equation}}
\setcounter{equation}{0}
\label{sec: Model ingredients}
In our chemical computations we examine how the steady state abundances of atomic and molecular hydrogen, H and H$_2$, and heavy metal-bearing species, especially OH, H$_2$O, O$_2$, CH and CO, depend on the overall heavy element abundances as parameterized by the metallicity scaling factor $Z'$, and on the ionization parameter $\zeta/n$. Here $\zeta$ is the total (primary plus secondary) H$_2$ ionization rate (s$^{-1}$), and $n \simeq n_{\rm H} + 2n_{\rm H_2}$ is the total volume density (cm$^{-3}$) of hydrogen nuclei in the gas parcels, where $n_{\rm H}$ and $n_{\rm H_2}$ are the atomic and molecular densities respectively. We assume that the parcels are exposed to steady sources of crx-ionization. We write the ionization rate as $\zeta=10^{-16} \zeta_{-16}$~s$^{-1}$ where $\zeta_{-16} \approx 1$ is the characteristic Galactic value as inferred via H$_3^+$ observations of clouds with H$_2$ column densities $\gtrsim 10^{21}$~cm$^{-2}$ (\citealt{McCall2003}, \citealt{Indriolo2012}, \citealt{Tielens2013}). We write the density $n=10^3 n_3$~cm$^{-3}$, where $n_3 \approx 1$ is the characteristic density for star-forming molecular clouds in the Milky Way \citep{McKee2007}. We adopt the chemical networks described in \S\ref{sec: Network}. We employ the reaction rate coefficients used by \citet{Boger2005a} which are based mainly on the UMIST99 database (\citealt{LeTeuff2000a}) with some updates. The data compiled in UMIST12 (\citealt{McElroy2013}) or KIDA (\citealt{Wakelam2012}) do not differ significantly for our reaction networks, and we have verified by explicit computation that our results are insensitive to the data set used\footnote{An exception is the position of the LIP/HIP boundary (see below) which can be sensitive to small rate-coefficient variations.}.

For the gas phase heavy element abundances, we adopt the \citet{Asplund2009} solar photospheric values (Table \ref{table: solar abundances}), multiplied by the overall metallicity factor $Z'$, with no grain depletion factors at any $Z'$. 
(We keep the helium abundance constant at a cosmological value of 0.1).
For H$_2$ formation on grains we assume a rate coefficient per hydrogen nucleus (\citealt{Hollenbach1971}; \citealt{Jura1974}; \citealt{Cazaux2002})
\begin{equation}
R=3 \times 10^{-17} \, T_2^{1/2} \, Z'^{\beta}\; \; \; \rm{cm}^3 \; \rm{s}^{-1} \; \; \; ,
\label{eq: grain rate}
\end{equation} where $T_2 \equiv T/100$~K.
The rate coefficient depends on the dust-to-gas ratio, and we assume that this varies as a power law, $Z'^{\beta}$, of the metallicity. 

\subsection{FUV off}
In the absence of externally incident FUV radiation, the steady state densities, $n_i$ (cm$^{-3}$), of the atomic and molecular species are determined by the set of formation-destruction rate equations,
\begin{align}
\sum_{jl} k_{ijl}(T) \, n_j \, n_l \, + \, \zeta \left[ \sum_{j} a_{ij}^{\rm D}  \, n_j\, + \, x_{\rm H_2} \sum_{j} a_{ij}^{\rm P} \,  n_j \right]  \nonumber\\
= n_i \left\{ \sum_{jl} k_{jil}(T) \, n_l \,  + \, \zeta \left[ \sum_{j} a_{ji}^{\rm D} \,  + \, x_{\rm H_2} \sum_{j} a_{ji}^{\rm P} \right] \right\} \; \; \; .
\label{eq: chemical equilibrium}
\end{align}
The $k_{ijl}(T)$ are the temperature dependent rate coefficients ($ \rm{cm}^{3} \ \rm{s}^{-1} $) for two-body reactions of species $j$ and $l$ that lead to the formation of $i$. The $a_{ij}$ are constants that multiply the total H$_2$ ionization rate $\zeta$, and are divided into two parts. 
The $a_{ij}^{\rm D}$ are for direct removal of species $j$ by the energetic (ionizing) particles, leading to the production of $i$. The $a_{ij}^{\rm P}$ are for induced photodestruction (crxp) by the internal UV photons produced by secondary-electron excitations of the H$_2$. The rates of the crxp processes are proportional to the H$_2$ gas density, and the $a_{ij}^{\rm P}$ are therefore multiplied by the H$_2$ fraction, $x_{\rm H_2} \equiv n_{\rm H_2}/n$.

The set of formation-destruction equations are augmented by mass and charge conservation. Thus,
\begin{equation}
\sum_i \, \alpha_{im} \, n_i \, = \, X_m \, n
  \label{eq: mass conservation}
\end{equation}
where $\alpha_{im}$ is the number of atoms of element $m$ contained in
species $i$, and $X_m$ is the total gas-phase abundance of element $m$ relative to the hydrogen density $n$ of nucleons. For the metals
\begin{equation}
\label{eq: X=AZ}
X_m \, = \, A_m \, Z' \; \; \; ,
\end{equation}
where $A_m$ is the solar abundance of element $m$ as given by Table \ref{table: solar abundances}. For helium we assume a constant cosmological abundance $A_{\rm He}=0.1$.
Charge conservation is,
\begin{equation}
\sum_i \, q_i \, n_i  \, = \, 0
\label{eq: charge conservation}
\end{equation}
where $q_i$ is the net charge of species $i$.

Dividing the rate equations by $n^2$ gives
 \begin{align}
\sum_{jl} k_{ijl}(T) \, x_j \, x_l \, + \,  \frac{\zeta}{n} \, \left[ \sum_{j} a_{ij}^{\rm D}  \, x_j\, + \, x_{\rm H_2} \sum_{j} a_{ij}^{\rm P} \,  x_j \right] \nonumber\\
= x_i \left\{ \sum_{jl} k_{jil}(T) \, x_l \,  + \, \frac{\zeta}{n} \left[ \sum_{j} a_{ji}^{\rm D} \,  + \, x_{\rm H_2} \sum_{j} a_{ji}^{\rm P} \right] \right\} \; \; \; ,
\label{eq: chemical equilibrium x}
\end{align} where $x_i \equiv n_i/n$ are the fractional abundances of species $i$ relative to the total hydrogen gas density. This shows that for a given $Z'$ (and with $X_m \, = \, A_m \, Z'$ for the metals) and for a given gas temperature, the fractional abundances depend only on the ionization parameter $\zeta/n$ (e.g., \citealt{Lepp1996a, Boger2005a}). We consider isothermal gas, and
thus the two basic parameters in our study are $\zeta/n$ and the metallicity $Z^{\prime}$.

The rate equations are non-linear, and we solve them iteratively using
Newton's method. 

\subsection{FUV on}
In the presence of externally incident FUV radiation (\S \ref{sub: Photoreactions}) photodissociation and photoionization processes must be added to the formation-destruction equations. These are then,
\begin{align}
&\sum_{jl} k_{ijl}(T) \, n_j \, n_l \, + \, \zeta \left[ \sum_{j} a_{ij}^{\rm D}  \, n_j\, + \, x_{\rm H_2} \sum_{j} a_{ij}^{\rm P} \,  n_j \right] \nonumber\\ 
&+ \, I_{\rm UV}\sum_{j} b_{ij} = n_i \left\{ \sum_{jl} k_{jil}(T) \, n_l \right. \nonumber\\
&+ \left. \, \zeta \left[ \sum_{j} a_{ji}^{\rm D} \,  + \, x_{\rm H_2} \sum_{j} a_{ji}^{\rm P} \right] 
\, + \, I_{\rm UV} \sum_{j} b_{ji} \right\} \; \; \; .
\label{eq: chemical equilibrium IUV on}
\end{align} 
 The $b_{ij} I_{\rm UV}$ in the additional terms are the photodissociation or photoionization rates ($ \rm{s}^{-1} $) of species $j$ that produce $i$ (Table \ref{table: photorates}). Dividing by $n^2$ we have,
\begin{align}
&\sum_{jl} k_{ijl}(T) \, x_j \, x_l \, + \,  \frac{\zeta}{n} \, \left[ \sum_{j} a_{ij}^{\rm D}  \, x_j\, + \, x_{\rm H_2} \sum_{j} a_{ij}^{\rm P} \,  x_j \right] \nonumber\\ 
&+ \,  \frac{I_{\rm UV}}{n} \sum_{j} b_{ij} \, x_j = x_i \left\{ \sum_{jl} k_{jil}(T) \, x_l \right. \nonumber\\   &+ \left. \, \frac{\zeta}{n} \left[ \sum_{j} a_{ji}^{\rm D} \,  + \, x_{\rm H_2} \sum_{j} a_{ji}^{\rm P} \right] \,  + \, \frac{I_{\rm UV}}{n} \sum_{j} b_{ji} \right\} \; \; \; .
\label{eq: chemical equilibrium x IUV on}
\end{align}
When FUV radiation is present a third parameter enters, $I_{\rm UV}/n$, the ratio of the FUV intensity to the gas density. Here $I_{\rm UV}$ refers to the local FUV intensity inside the gas parcel, after accounting for any shielding by an outer dust-absorption layer. 

We assume that the LW band is always fully blocked by a shielding H$_2$ gas column of at least $10^{22}$~cm$^{-2}$. For such shielding columns FUV photodissociation is always negligible compared to crx-ionization of the H$_2$
in the gas parcels.
To keep the cloud sizes reasonably small we assume that
the column densities of any photodissociated atomic (H{\small I}) envelopes are 
negligible compared to the H$_2$ shielding columns, even at very low $Z'$.  
This requires that 
\begin{equation}
\label{eq: min shielding}
n \ \gtrsim \ 2 \times\frac{I_{\rm UV}^0}{Z'^{\beta}} \ \ \ {\rm cm}^{-3} \ ,
\end{equation}
where $I_{\rm UV}^0$ is the free-space FUV radiation intensity factor.
This follows from the \citet{Sternberg2014} expression for the photodissociated H{\small I} column in the ``weak field" limit ($I^0_{\rm UV}/n_3\lesssim10$), 
and given by $N_{\rm HI} \approx 0.1F_0/(2Rn) = 4 \times 10^{19} Z'^{-\beta} \,  (I^0_{\rm UV}/n_3)$~cm$^{-2}$. In this expression, 
$F_0=2.07\times10^7$~cm$^{-2}$ is the characteristic
photodissociation photon flux for $I^0_{\rm UV}=1$ (\citealt{Draine2003}; \citealt{Sternberg2014}). For $\beta=1$ and $I^0_{\rm UV}=1$
Equation (\ref{eq: min shielding}) requires 
$n\gtrsim 2\times10^3$~cm$^{-3}$ for $Z'=10^{-3}$.  
The corresponding
cloud sizes including the shielding envelopes are then $\lesssim 10$~pc.

For an LW H$_2$ blocking column $N_{\rm H_2}^{\rm block}$=10$^{22}$~cm$^{-2}$ the associated 1000~\AA$\,$ dust opacity is $\tau_g = 2 \sigma_g N_{\rm H_2}^{\rm block} \approx 38 Z'^{\beta}$, where $\sigma_g = 1.9 \times 10^{-21}  Z'^{\beta}$~cm$^2$ is the dust cross-section per hydrogen nucleon.
Thus when turning on the FUV field in Equations (\ref{eq: chemical equilibrium IUV on}) and (\ref{eq: chemical equilibrium x IUV on}), we set
\begin{equation}
\label{eq: IUV IUV0}
I_{\rm UV} \, = \, I_{\rm UV}^0 \, \mathrm{e}^{-\tau_g}  \, = \, I_{\rm UV}^0 \, \mathrm{e}^{-38Z'^{\beta}} \; \; \; ,
\end{equation}
In this expression we are assuming that the dust opacity associated with any H{\small I} in the shielding layer is negligible.

Our expression for $I_{\rm UV}$ allows the local field intensity inside the parcels to vary smoothly as we vary $Z'$, and as the effects of dust shielding are altered. 
For simplicity, and because we are assuming an arbitrary shielding column, we ignore the wavelength dependence of the dust attenuation for the individual species in our set.
For high enough $Z'$, the dust opacity becomes large and all of the photorates vanish due to significant dust shielding. When the metallicity is low, $\tau_g$ is small, but the LW band remains blocked. Species that are removed by $\lambda > 1108$~\AA $\,$ photons are only partially shielded, or not at all for sufficiently small $Z'$.

\section{H/H$_2$ Balance Analytic Treatment, and Time Scales}
\label{sec: H_H2_analytical}
Before presenting our detailed chemical computations (in \S\ref{sec: Two illustrative} and \S\ref{sec: parameter space}) we consider just H$_2$ formation-destruction and the behavior of the H/H$_2$ density ratio, in a simplified analytic treatment, showing the dependence on $\zeta/n$ and $Z'$. Our analysis generalizes previous discussions \citep[e.g.,][]{deJong1972, Glover2003, Cazaux2004}. 
We also consider the dependences of the H$_2$ formation time-scales on $Z'$ and $\zeta/n$. The H$_2$ formation time-scales determine the overall conditions required for chemical equilibrium.

\subsection{Steady State H/H$_2$}
\label{sub: Steady State}
 As discussed in \S\ref{sub: H and He} the two primary H$_2$ formation channels are grain catalysis (reaction [\ref{R: H2 form dust}]) and gas phase production via H$^-$ ([\ref{R: H- form}] and [\ref{R: H2 form gas}]). In the absence of FUV photodissociation, the H$_2$ is removed by crx-ionization (reactions [\ref{R: H2 dist cr form H+}] and [\ref{R: H2 dist cr form H2+}]) leading mainly to the formation of H$_2^+$. When the gas is primarily molecular, the destruction rate is {\it enhanced} by further reactions of the H$_2$ with H$_2^+$ (reaction [\ref{R: H3+ form}]). However, when the gas is atomic the net H$_2$ removal rate is {\it reduced} by electron charge transfer from H to H$_2^+$, leading back to H$_2$ (reaction [\ref{R: H2 form by H2+}]).

  \begin{figure}
 \includegraphics[width=.5\textwidth]{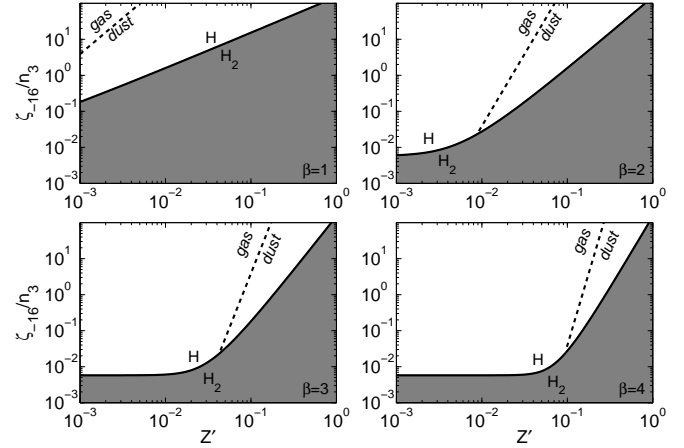}
\caption{Atomic (H) and molecular (H$_2$) regimes in the $\zeta_{-16}/n_3$ versus $Z'$ parameter space, assuming $R=R_0 Z'^{\beta}$, for $\beta$=1, 2, 3 and 4 (see text). The solid curves are the H-to-H$_2$ transition lines where $n_{\rm H}=n_{\rm H_2}$. The hydrogen is molecular below the solid curves (shaded) and atomic above them. In the atomic regimes, the dashed lines separate the zones where gas-phase versus dust catalysis dominate the H$_2$ formation.}

 \label{fig: H2 zones}
 \end{figure}
 
Thus, the H/H$_2$ formation-destruction equation for steady state conditions is
\begin{equation}
 \label{eq: H2/H not devided by n2}
\zeta \times y \times f_d \times n_{\rm H_2} \, = \, [ R \, n \,  + \, k_2 \, n_e] \, n_{\rm H}  \; \; \; ,
 \end{equation} with
 \begin{equation}
 \label{eq: H2 conservation not devided by n2}
 2 \, n_{\rm H_2} \, + \, n_{\rm H} \, = \, n  \; \; \; ,
 \end{equation} or
 \begin{equation}
 \label{eq: H2/H}
\left( \frac{\zeta}{n} \right) \times y \times f_d \times x_{\rm H_2} \, = \, [ R \,   + \, k_2 \, x_e] \, x_{\rm H}  \; \; \; ,
 \end{equation} with
  \begin{equation}
 \label{eq: H2 conservation}
 2 \, x_{\rm H_2} \, + \, x_{\rm H} \, = \, 1  \; \; \; .
 \end{equation} 
Here $n_{\rm H}$ and $n_{\rm H_2}$, and $x_{\rm H}$ and $x_{\rm H_2}$, are the atomic and molecular hydrogen densities and fractions, $\zeta$ is the crx-ionization rate, and $R$ is the grain-surface H$_2$ formation rate coefficient. The rate coefficient $k_2 = 2.0 \times 10^{-16} \, T_2^{0.67}$~cm$^3$~s$^{-1}$ (independent of $Z'$) is for radiative association of hydrogen atoms with electrons (reaction [\ref{R: H- form}]) which is the rate limiting step for gas-phase H$_2$ formation. On the left-hand side, the factor $y \simeq 2$ in the molecular regime ($2 x_{\rm H_2} \simeq 1$) where the H$_2$ destruction rate is enhanced by the further reactions with H$_2^+$. In the atomic regime ($x_{\rm H} \simeq 1$) the parameter $y \simeq 1$. The factor $f_d$ is the fraction
 \begin{equation}
 \label{eq: H2 destruction branching ratio}
 f_d \, \equiv \, \frac{k_{\ref{R: H3+ form}} \, x_{\rm H_2} } {k_{\ref{R: H3+ form}} \, x_{\rm H_2} \, + \, k_{\ref{R: H2 form by H2+}} \, x_{\rm H} } \; \; \; ,
 \end{equation} 
of all H$_2$ ionizations that are not followed by charge-transfer back to H$_2$. In the molecular regime $f_d \simeq 1$. In the atomic regime $f_d \simeq 3.3 x_{\rm H_2}/x_{\rm H}$.
 
In this analysis we assume that the dust-to-gas ratio varies as a power $\beta$ of the metallicity as given by Equation (\ref{eq: grain rate}), and write
\begin{equation}
\label{eq: R R0}
R \, = \, R_0 \, Z'^{\beta} \; \; \; ,
\end{equation}
where $R_0 = 3 \times 10^{-17} \, T_2^{1/2}$~cm$^3$~s$^{-1}$ is the rate coefficient for solar ($Z'$=1) metallicity. (In our full chemical computations in \S\ref{sec: Two illustrative} and \S\ref{sec: parameter space} we set $\beta=1$).


In Equation (\ref{eq: H2/H}), the gas phase formation term, $k_2 x_{\rm e}$, can become important only when the hydrogen is primarily atomic since then a relatively large H$^-$ abundance can be maintained by electron attachment. We estimate $x_{\rm e}$ in the atomic-regime via the condition of ionization-recombination equilibrium
\begin{equation}
0.46 \zeta \, n_{\rm H} \, \simeq \, 0.46 \zeta \, n \, = \, \alpha_B \, n_{\rm H^+} \, n_{\rm e} \, \simeq \, \alpha_B \, n_{\rm e}^2 \; \; \; ,
\end{equation} or
\begin{align}
x_{\rm e} \, \equiv \, \frac{n_{\rm e}}{n} \, &\simeq \, 0.68 \, \left( \frac{\zeta}{\alpha_B n} \right)^{1/2} 
\nonumber\\
\, &= \, 7.6 \times 10 ^{-5} \, T_2^{0.38}  \,\left(\frac{\zeta_{-16}}{n_3} \right)^{1/2} \; \; \; .
\label{eq: xe}
\end{align} Here we have assumed $\alpha_B=8.0 \times 10^{-12} T_2^{-0.75}$~cm$^{3}$~s$^{-1}$ for case B radiative recombination.

Given expressions (\ref{eq: R R0}) and (\ref{eq: xe}) for $R$ and $x_{\rm e}$, and setting $x_{\rm H}=x_{\rm H_2}$ in Equation (\ref{eq: H2/H}), we obtain the relation between $\zeta/n$ and $Z'$ for which the atomic and molecular densities are equal. These are the solid curves in Figure \ref{fig: H2 zones}, for $\beta$=1, 2, 3, and 4, and assuming $T_2=1$.

For any $\beta$, the curves flatten when $Z'$ is sufficiently small and gas-phase H$_2$ formation dominates. In this limit $Rn$ is negligible in Equation (\ref{eq: H2/H}), and for $x_{\rm H}=x_{\rm H_2}$
\begin{equation}
f_d \times y \times \frac{\zeta}{n} \, = \, k_2 \, x_e \; \; \; .
\end{equation}
Assuming $y=2$, $f_d=1$ at the H-to-H$_2$ transition point (justified by the results of our detailed numerical solutions), and assuming Equation (\ref{eq: xe}) for the fractional ionization, the ionization parameter for which $x_{\rm H}=x_{\rm H_2}$ is given by
\begin{equation}
\frac{\zeta}{n} \, \simeq \, 0.12 \, \frac{k_2^2}{\alpha_B} \; \; \; .
\end{equation}
Thus for gas-phase H$_2$ formation (low $Z'$ limit), the atomic-to-molecular transition occurs at
\begin{equation}
\frac{\zeta_{-16}}{n_3} \simeq 5.7 \times 10^{-3} \, T_2^{2.1} \; \; \; ,
\label{eq: transition line gas} 
\end{equation}
as seen in Figure \ref{fig: H2 zones} for very low $Z'$.

At sufficiently high $Z'$, grain H$_2$ catalysis dominates, the H$_2$ formation efficiency increases as $Z'^{\beta}$, and the H/H$_2$ transition curves turn upward. Then, neglecting the gas phase formation term $k_2 n_{\rm e}$ in Equation (\ref{eq: H2/H}), and again setting $y=2$ and $f_d=1$ we have
\begin{equation}
\frac{\zeta}{n} \, \simeq \, 0.5 \, R_0 \, Z'^{\beta} \; \; \; .
\end{equation} This gives
\begin{equation}
\frac{\zeta_{-16}}{n_3} \, \simeq \, 1.5 \times 10^{2} \, T_2^{1/2} \, Z'^{\beta} 
\label{eq: transition line dust}
\end{equation} for the H/H$_2$ transition curves for H$_2$ grain catalysis. For example, for $\beta=2$, the H-to-H$_2$ transition occurs at $\zeta_{-16}/n_3 = 1.5 T_2^{1/2}$ for $Z'=0.1$, or at $\zeta_{-16}/n_3 = 1.5 \times 10^2 T_2^{1/2}$ for $Z'=1$.

In Figure \ref{fig: H2 zones}, the dashed lines delineate the zones where  gas-phase production (to the left) versus grain catalysis (to the right) dominates H$_2$ formation in the atomic regimes ($x_{\rm H} \simeq 1$) above the solid curves. We draw these by equating the two formation rates
\begin{equation}
R_0 \, Z'^{\beta} \, = \, x_{\rm e} k_2 \; \; \; 
\end{equation}
and using Equation (\ref{eq: xe}) for $x_{\rm e}$. This gives
\begin{equation}
\frac{\zeta}{n} \, = \, 2.2 \, \left( \frac{R_0 \, Z'^{\beta}}{k_2} \right)^2 \, \alpha_B \; \; \; ,
\end{equation} or
\begin{equation}
\frac{\zeta_{-16}}{n_3} = 4.0 \times 10^6 \; T_2^{-1.1} \, Z'^{2\beta} 
\label{eq: transition between gas and dust} \; \; \; ,
\end{equation}
for the dashed lines in Figure \ref{fig: H2 zones}. For example, for $\beta=2$ the gas phase and grain H$_2$ formation rates are equal for $\zeta_{-16}/n_3 = 4 \times 10^{-2} T_2^{-1.1}$ for $Z'=10^{-2}$, or $\zeta_{-16}/n_3 = 4 \times 10^2 T_2^{-1.1}$ for $Z' = 0.1$.

We may now define the critical metallicity $Z'_{\rm crit}$ at which the gas-phase and dust-grain formation rates are equal at the H-to-H$_2$ transition point. For $Z' < Z'_{\rm crit}$ the H-to-H$_2$ transition is controlled by the gas-phase formation. For $Z' > Z'_{\rm crit}$ dust-formation dominates the transition. The critical metallicities occur at the intersections of the dashed and solid curves in Figure \ref{fig: H2 zones}. These intersections may be estimated by equating Equations (\ref{eq: transition line gas}) and (\ref{eq: transition line dust}). This gives
\begin{equation}
\label{eq: Zcrit}
 Z'^{\beta}_{\rm crit} \, \simeq \, 0.23 \, \frac{k_2^2}{2 \times \alpha_B \, R_0} \, = \, 3.8 \times 10^{-5} \; T_2^{1.6} \; \; \;.
\end{equation} 
For $\beta$ increasing from 1 to 4, $Z'_{\rm crit}$ ranges from $4 \times 10^{-5} T_2^{1.1}$ to $8 \times 10^{-2} T_2^{1.6}$. 

Finally, expressions for the H$_2$ fraction in the atomic regime ($x_{\rm H} \simeq 1$) may now be written down.
For gas-phase formation, Equation (\ref{eq: H2/H}) takes the form
\begin{equation}
\frac{x_{\rm H_2}}{x_{\rm H}} \, \simeq  \, k_2 \, x_{\rm e} \, \frac{1}{f_d} \, \frac{1}{y} \, \left( \frac{\zeta}{n} \right)^{-1} \; \; \; .
\end{equation} Setting $y=1$ and $f_d \simeq 3.25 x_{\rm H_2}/x_{\rm H}$ as appropriate for the atomic regime, and using Equation (\ref{eq: xe}) for the electron fraction, we obtain 
\begin{equation}
\label{eq: xH2 gas}
x_{\rm H_2} \, \simeq \,  \frac{x_{\rm H_2}}{x_{\rm H}} \, \simeq 
\ 0.22 \, T_2^{0.5} \, \left(\frac{\zeta_{-16}}{n_3} \right)^{-1/4}  \; \; \; .
\end{equation} 
For dust catalysis
\begin{equation}
\frac{x_{\rm H_2}}{x_{\rm H}} \, \simeq \, R_0 \, Z'^{\beta} \, \frac{1}{f_d} \, \frac{1}{y} \; \; \; ,
\end{equation} and
\begin{equation}
\label{eq: xH2 dust}
x_{\rm H_2} \, \simeq \, \frac{x_{\rm H_2}}{x_{\rm H}} \, 
\simeq \, 0.95 \,  T_2^{1/4} \,  \left(\frac{\zeta_{-16}}{n_3} \right)^{-1/2} \, \left(\frac{Z'}{10^{-2}} \right)^{\beta/2}  \; \; \; .
\end{equation} 
Where again we set $y=1$ and $f_d \simeq 3.25 x_{\rm H_2}/x_{\rm H}$ as appropriate for the atomic regime.


\subsection{Time Scales}
\label{sub: time scales}
In the analysis above and in our computations in \S\ref{sec: Two illustrative} and \S\ref{sec: parameter space} we are assuming that the systems are in a steady state, and that the chemical equilibrium times are shorter than the
cloud lifetimes. Overall equilibrium is set by the relatively long time-scale for H$_2$ formation. For grain catalysis the H$_2$ formation time scale is
\begin{equation}
\label{eq: time scale grain}
t_{\rm H_2}^{\rm dust} \, = \, \frac{1}{R \, n} \, \approx \, 10^6 \ T_2^{-1/2} n_3^{-1} Z'^{-\beta}  \; \; \; \rm{yr} \; \; \; .
\end{equation}
For the gas phase formation via H$^-$ the time scale is
\begin{equation}
\label{eq: time scale gas}
t_{\rm H_2}^{\rm gas} \, = \, \frac{1}{k_2 \, n \, x_{\rm e}} \, \approx \, 2 \times 10^9 \ T_2^{-1.1} \ n_3^{-1/2} \ \zeta_{-16}^{-1/2}   \; \; \; \rm{yr} \; \; \; ,
\end{equation} where we used Equation (\ref{eq: xe}) for $x_{\rm e}$. The ratio of these time scales is
\begin{equation}
\label{eq: time scales ratio}
\frac{t_{\rm H_2}^{\rm dust}}{t_{\rm H_2}^{\rm gas}} \ = \ \frac{k_2 \, n \, x_{\rm e}}{R \, n} \, \approx \, 5 \times 10^{-4} \ T_2^{0.6} \,  \left(\frac{\zeta_{-16}}{n_3} \right)^{1/2} \, \frac{1}{Z'^{\beta}}  \; \; \; .
\end{equation}
At solar ($Z'=1$) metallicity, H$_2$ formation on dust is much faster than in the gas-phase unless the gas density is unusually low, or the ionization parameter is extremely large. At low $Z'$ and low dust-to-gas ratios, the gas phase formation route may become faster even at characteristic densities.
Both time-scales are generally short compared to the present day age of the universe (13.8~Gyr), and the typical ages of galaxies, and H$_2$ formation is rapid for shielded gas. However, at the early reionization epoch ($z \approx 10$, or 0.48~Gyr, \citealt{Hinshaw2013}), the H$_2$ formation time-scale may become long
irrespective of the cloud evolution time scales.
For example, at reionization, and an early metal abundance $Z'$=10$^{-3}$ 
but for a Galactic $\zeta_{-16}/n_3=1$, the dust and gas-phase H$_2$ formation time-scales are comparable even for $\beta=1$. A large cloud density,
$n \gtrsim 2 \times 10^{3}$~cm$^{-3}$, is then required to achieve a chemical steady state, for complete conversion of H to H$_2$ within the corresponding Hubble time.

\section{ Analysis and Scaling Relations }
\label{sec: Two illustrative}
With the above analytic results for the steady-state H/H$_2$ balance and time scales, we now show full chemical computations for several illustrative model sequences, as one-dimensional (1D) cuts through our parameter space. We first present models for varying $Z'$ at fixed $\zeta/n$, and second for varying $\zeta/n$ at fixed $Z'$. In all of these computations we assume that the H$_2$ formation rate coefficient varies linearly with $Z'$ (i.e., we set $\beta$=1 in Equation [\ref{eq: grain rate}]), and solve Equations (\ref{eq: mass conservation})-(\ref{eq: chemical equilibrium x}) and (\ref{eq: chemical equilibrium x IUV on}) for our chemical networks, without and with FUV. For FUV on we assume the photorates for the diluted 10$^5$~K radiation field (see Table \ref{table: photorates}).

\subsection{Dependence on $Z'$ for fixed $\zeta/n$.}
\label{sub: Z dependence}
\begin{figure}
 \includegraphics[width=.5\textwidth]{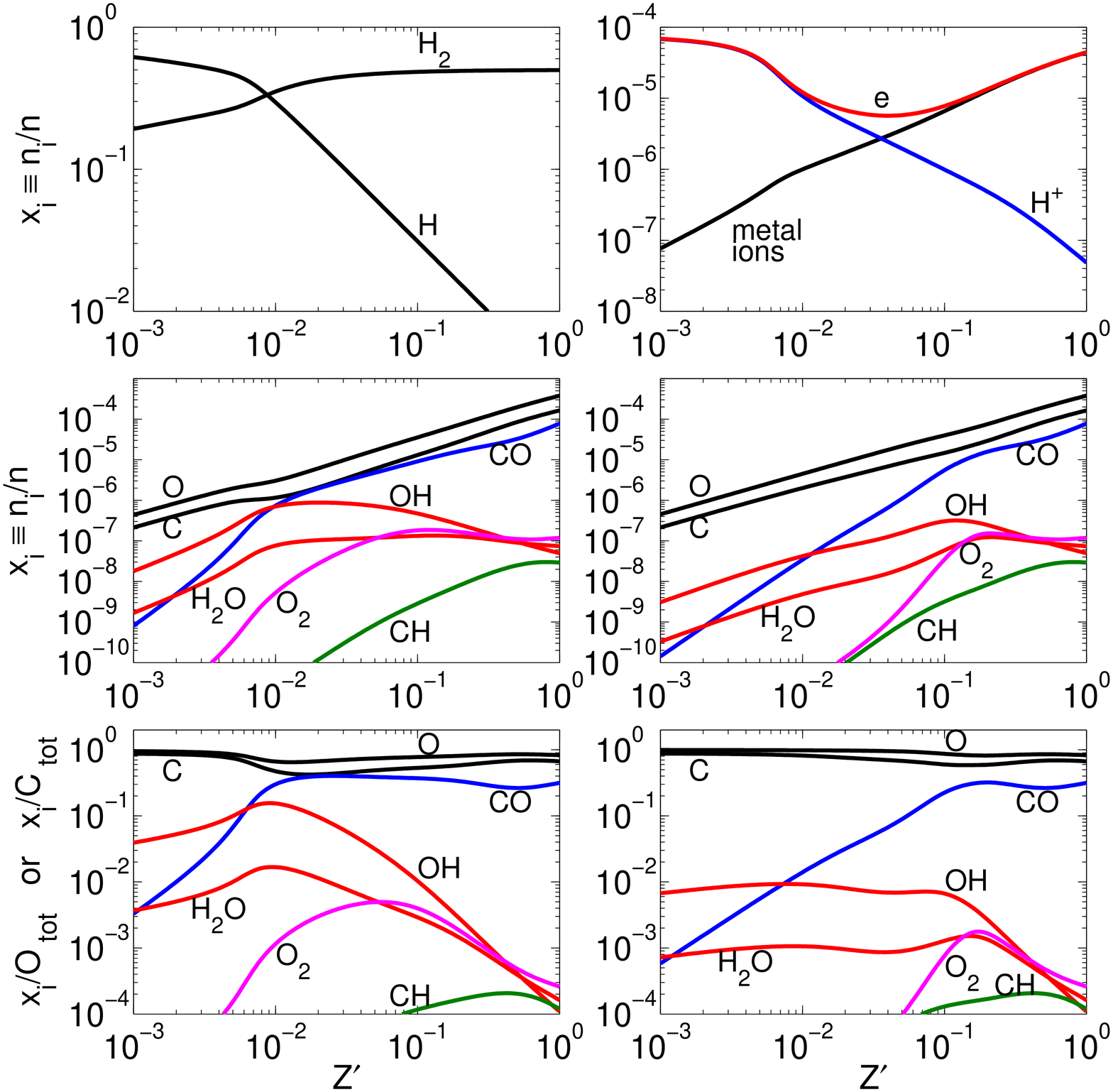}
\caption{Abundances versus metallicity for $\zeta_{-16}/n_3$=1 and $T_2$=1. 
The upper and middle panels show abundances relative to total hydrogen nuclei, $x_i \equiv n_{\rm{i}}/n$ ($n=n_{\rm H} + 2 n_{\rm H_2}$). In the bottom panels we normalize to the total elemental carbon and oxygen abundances (C$_{\rm tot} = 2.9 \times 10^{-4} Z'$ and O$_{\rm tot} = 4.9 \times 10^{-4} Z'$). In the lower right panel $I_{\rm UV}^0/n_3$=1, in all other panels $I_{\rm UV}^0/n_3$=0.}
 \label{fig: x_vs_Z}
 \end{figure}

In Figure \ref{fig: x_vs_Z} we display the steady-state abundances, $x_i\equiv n_i/n$, for H, H$_2$, H$^+$, e, C, O, OH, H$_2$O, O$_2$, CH and CO. In these computations we set $\zeta_{-16}/n_3=1$, and vary $Z'$ from 1 to 10$^{-3}$. Again, we are assuming that the LW band is always blocked so that H$_2$ and CO photodissociation is always excluded whether or not an FUV background field is present. 

The upper left panel shows the behavior for H and H$_2$.
For $Z' \gtrsim 10^{-2}$, H$_2$ formation by dust-grain catalysis
is rapid compared to H$_2$ removal by crx-ionization, and the gas is fully molecular. The atomic hydrogen fraction increases as grain H$_2$ formation becomes less efficient with decreasing $Z'$ according to Equation (\ref{eq: grain rate}). The H-to-H$_2$ transition occurs at $Z' \approx 10^{-2}$, and this is consistent with Equation (\ref{eq: transition line dust}) for $\zeta_{-16}/n_3$=1. The H$_2$ abundance continues to drop as $Z'$ is reduced and as the molecular formation efficiency decreases. 
The curve for \H{2} finally flattens as $Z'$ becomes very small and gas-phase formation dominates, however this occurs only for $Z'<10^{-3}$ below the minimum $Z'$ that we consider. 




The upper-right panel of Figure \ref{fig: x_vs_Z} shows the electron fraction $x_{\rm e}$. It also shows the summed fractions of the metal positive charge carriers. For our assumed $\zz=1$, these are mainly the atomic ions C$^+$, Si$^+$, and S$^+$. The metal positive charge carriers are collectively labeled ``metal-ions" in Figure \ref{fig: x_vs_Z}. This panel also shows the proton fraction \Hp{}.
In general, the interstellar gas-phase chemical sequences yield two types of solutions (HIP and LIP) for the equilibrium ionization states, depending on $\zeta/n$ and the gas-phase
elemental abundances \citep{Oppenheimer1974, PineaudesForets1992, LeBourlot1993, Lee1998, Boger2006a, Wakelam2006}.
In the ``high-ionization-phase" (HIP) the electrons are removed by a combination of H$_3^+$ dissociative recombination and radiative recombination of the atomic metal ions. In the ``low-ionization-phase" (LIP) the electrons are removed mainly by dissociative recombination with molecular metal ions. In purely gas-phase systems the HIP to LIP transition can be abrupt due to instabilities in the chemical networks (e.g.~\citealt{Boger2006a}). The model sequence in Figure \ref{fig: x_vs_Z} is all HIP for the full range of $Z'$ for our assumed $\zeta_{-16}/n_3$=1.

At $Z'$=1, $x_e \simeq 5 \times 10^{-5}$ and the positive charge is carried by the metal ions.
As $Z'$ is reduced the total metal ion fraction drops linearly, and $x_{\rm e}$ decreases accordingly. Meanwhile, the proton fraction rises with the increasing abundance of free hydrogen atoms available for direct crx-ionization (reaction [\ref{R: H dist form H+}]). The decreasing metal-ion densities together with the increasing abundance of protons leads to the minimum $x_{\rm e} \simeq 6 \times 10^{-6}$ at $Z'\approx 4\times 10^{-2}$. At very low $Z'$, H$^+$ becomes the dominant positive charge carrier, and $x_{\rm e}$ increases until the gas becomes fully atomic. He$^+$ does not contribute significantly because it is rapidly removed by charge transfer (reactions [\ref{R: He+ dist High Z}]-[\ref{R: He dist H2 to He H+ H}]).
In the fully atomic regime the electron fraction reaches a plateau of $x_{\rm e} \approx 7 \times 10^{-5}$, consistent with Equation (\ref{eq: xe}) for the balance between hydrogen ionization and electron-proton radiative recombination.

The middle-left panels of Figure \ref{fig: x_vs_Z} show the abundance fractions $x_i$ versus $Z'$, for C, O, OH, H$_2$O, O$_2$, CH, and CO, for FUV off. These fractions are relative to the total hydrogen gas density.
In the lower-left panels we normalize relative to the total available oxygen or carbon at each $Z'$, and plot curves for C/C$_{\rm tot}$, CH/C$_{\rm tot}$, CO/C$_{\rm tot}$, O/O$_{\rm tot}$, OH/O$_{\rm tot}$, H$_2$O/O$_{\rm tot}$, and O$_2$/O$_{\rm tot}$, (where O$_{\rm tot} \equiv A_{\rm O} Z' = 4.9 \times10^{-4} Z'$ and C$_{\rm tot} \equiv A_{\rm C} Z' = 2.9 \times10^{-4} Z'$).

Several important features and trends can be seen. First, at high $Z'$, the dominant metal-bearing molecule is CO. For example for $Z' \approx 1$, $x_{\rm CO} \approx 8 \times 10^{-5}$ and CO/C$_{\rm tot} \approx 0.3$. When the hydrogen is molecular ($Z' \gtrsim 10^{-2}$), at least 30\% of the available carbon is locked in CO with the remaining carbon in atomic form.
Because O$_{\rm tot}$/C$_{\rm tot} >1$ in our models 
a large fraction of the oxygen always remains atomic. The large CO abundance is the familiar chemical state for standard interstellar molecular clouds. Figure \ref{fig: x_vs_Z} shows that CO production is efficient at all $Z'$ so long as the hydrogen is molecular. In the absence of selective photodissociation the CO then serves as a good proxy for the H$_2$ (e.g. \citealt{Bolatto2013}). We refer to this limit as the ``CO-dominated regime".
When the hydrogen is atomic ($Z' \lesssim 10^{-2}$ for $\zz=1$) the CO abundance drops sharply and then OH becomes the most abundant molecule containing a heavy element. We refer to this limit as the ``OH-dominated" regime. Figure \ref{fig: x_vs_Z} shows that the transition from the OH to CO dominated regimes occurs close to the H-to-H$_2$ transition point. As we discuss further below this is  ly true for any $\zeta/n$. A large atomic carbon abundance is generally maintained when \z is sufficiently large for a given $Z'$, and the C/CO density ratio is then of order of unity. At low $\zeta/n$ the C/CO ratio always becomes small. In Figure \ref{fig: x_vs_Z} we are in the large \z regime.

Second, the OH, H$_2$O, O$_2$, and CH abundances are all comparable for large $Z'$. For our assumed $\zeta_{-16}/n_3$=1, $x_{\rm OH}$, $x_{\rm H_2O}$, $x_{\rm O_2}$, and $x_{\rm CH}$ are $\sim 10^{-8}$ at $Z'=1$, as seen in Figure \ref{fig: x_vs_Z}. However, the abundance curves for these four species diverge sharply as $Z'$ becomes small.

The behavior for OH is of particular interest. 
As $Z'$ is reduced, the OH abundance first {\it increases}, and reaches a maximum near the H/H$_2$ transition. At this point, $x_{\rm OH} = 6 \times 10^{-7}$, and OH/O$_{\rm tot} = 0.16$, and a significant fraction of the gas phase oxygen is locked in OH molecules. In particular, the OH/CO abundance ratio $\sim 1$ at this point. At still lower $Z'$ the OH abundance decreases, but once the hydrogen becomes predominantly atomic OH/O$_{\rm tot}$ approaches a constant value of 2\%. 
In contrast, CO/C$_{\rm tot}$ decreases linearly with $Z'$, and OH/CO becomes large.

The variation of $x_{\rm OH}$ with $Z'$ and $\zeta/n$ is essential and can be understood via an approximate analytic scaling relation, as follows. (We show numerical results for \OH versus $\zeta/n$ at fixed $Z'$ in \S\ref{sub: zetan dependence}). At high $Z^\prime$, and in the molecular regime ($2 x_{\rm H_2} \simeq 1$), 
crx-ionization of H$_2$ leads to the formation of H$^+$ or H$_3^+$ that then react with atomic oxygen (reactions [\ref{R: O+ form}] and [\ref{R: OH+ form with H3+}]) initiating the abstraction sequence leading to OH. 
At temperatures $\gtrsim 50$~K the  
H$^+$ density is limited by charge transfer with atomic oxygen
(reaction [\ref{R: O+ form}]).
The OH formation rate via H$^+$ 
is then equal to the dissociative H$_2$ ionization rate, and independent of metallicity. 
For the HIP conditions in Figure \ref{fig: x_vs_Z} the H$_3^+$ density is limited by dissociative recombination ([\ref{R: H3+ recombiantion 3H}] and [\ref{R: H3+ recombination H2 H}]). 
The relative rates of the H$^+$ versus H$_3^+$ routes therefore depends on the atomic oxygen to electron ratio.
The OH is removed mainly by atomic oxygen (reaction [\ref{R: OH dist high Z with O}]) hence the removal rate is proportional to $Z'$, and independent of $\zeta/n$.
To a good approximation we then have
\begin{align}
\label{eq: OH highZ}
x_{\rm OH} \, &\propto \, \left( \frac{\zeta}{n} \right) \, \left[ 1 \, + \frac{0.93 k_{\rm \ref{R: OH+ form with H3+}}}{0.02 (k_{\rm \ref{R: H3+ recombiantion 3H}}+k_{\rm \ref{R: H3+ recombination H2 H}})} \, \frac{x_{\rm O}}{x_{\rm e} } \right] \frac{1}{Z'} \nonumber\\
&= \, \left( \frac{\zeta}{n} \right) \, \left[ 1 \, + 0.27 \, \frac{x_{\rm O}}{x_{\rm e} } \right] \frac{1}{Z'} \; \; \; ,
\end{align} 
where $k_{\rm \ref{R: OH+ form with H3+}}$, $k_{\rm \ref{R: H3+ recombiantion 3H}}$, $k_{\rm \ref{R: H3+ recombination H2 H}}$ are the rate coefficients for reactions [\ref{R: OH+ form with H3+}], [\ref{R: H3+ recombiantion 3H}] and [\ref{R: H3+ recombination H2 H}] (evaluated for $T_2=1$), and the terms in brackets account for the H$^+$ and H$_3^+$ formation channels respectively. To first-order, $x_{\rm OH}$ is proportional to $\zeta/n$, and for a given hydrogen gas density the OH abundance is a measure of the ionization rate (e.g., \citealt{Black1973, VanDishoeck1986}). 
In Equation (\ref{eq: OH highZ}),
the H$^+$ formation channel dominates when the oxygen-to-electron ratio, $x_{\rm O}/x_{\rm e}$, is small. The oxygen-to-electron ratio is independent of $Z'$ when the positive charge is carried by metal species as is the case for high $Z'$, but it does vary inversely with $\zeta/n$ and this moderates the sensitivity of $x_{\rm OH}$ to $\zeta/n$ (as discussed further in \S\ref{sub: zetan dependence}). 
To first order the OH formation rate is independent of metallicity.
The 1/$Z'$ term in Equation (\ref{eq: OH highZ}) accounts for the removal of the OH by atomic oxygen. Figure \ref{fig: x_vs_Z} shows the approximately linear rise in $x_{\rm OH}$ as the metallicity is reduced.

In the opposite limit; low $Z'$, atomic regime ($x_{\rm H} \simeq 1$), and for gas-phase H$_2$ production, OH formation proceeds via reactions of O$^+$ ions with the trace H$_2$ available (reaction [\ref{R: OH+ form}]). But now the O$^+$ density
is set by charge transfer equilibrium with atomic hydrogen (reaction [\ref{R: O+ form}]),
and the H$^+$ is removed by electron recombination. Thus, in this limit the
OH formation rate is linearly proportional to the oxygen abundance and metallicity.
Removal of the OH is via reactions with H$^+$ [\ref{R: OH dist lowZ}] rather then by 
oxygen and other metals. In this limit
\begin{equation}
\label{eq: OH lowZ without zetan dependence}
x_{\rm OH} \, \propto \, x_{\rm H_2} \, \frac{1}{f} \, \frac{  \, x_{\rm O^+} }{x_{\rm H^+} } \, \propto \, x_{\rm H_2} \, \frac{1}{f} \, \frac{  \, x_{\rm O} }{x_{\rm H} } \, \propto \, x_{\rm H_2} \, \frac{1}{f} \, Z' \; \; \; .
\end{equation} 
Here $f \propto x_{\rm e}/x_{\rm H_2}$ is the fraction of OH$^+$ productions (via [\ref{R: OH dist lowZ}]) that do not loop back to OH via the abstraction sequence [\ref{R: H2O+ form}]-[\ref{R: OH form 2 (OH,H2)}]. In Equation (\ref{eq: OH lowZ without zetan dependence}), $x_{\rm O^+}/x_{\rm H^+} \propto x_{\rm O}/x_{\rm H}$ as set by forward and backward charge-transfer equilibrium. Furthermore, from Equations (\ref{eq: H2/H}) and (\ref{eq: xe}), we have $x_{\rm e} \propto (\zeta/n)^{1/2}$ and $x_{\rm H_2} \propto (\zeta/n)^{-1/4}$. Thus at low $Z'$, $x_{\rm OH}$ is proportional to $Z'$, and 
\begin{equation}
\label{eq: OH lowZ Z and zetan dependence}
x_{\rm OH} \, \simeq \, 0.02 \, A_{\rm O} Z' \, \left(\frac{\zeta_{-16}}{n_3} \right)^{-1} \; \; \; ,
\end{equation} where $A_{\rm O} = 4.9 \times 10^{-4}$ is the oxygen abundance at solar metallicity.
In expression (\ref{eq: OH lowZ Z and zetan dependence}) the prefactor of 2\% is obtained from our numerical results, as shown in Figure \ref{fig: x_vs_Z}. Importantly, the OH fraction varies {\it inversely} with $\zeta/n$ in the low $Z'$ limit. This is due to (a) the reduced H$_2$ fraction with increasing ionization parameter, 
and (b) the increased electron to H$_2$ ratio that enhances the efficiency of OH destruction by protons.

The behavior for H$_2$O and O$_2$ 
differs due to the variety of removal processes with metal atoms and ions, [\ref{R: H2O dist with C+}]-[\ref{R: H2O dist with H3+}] for H$_2$O, and [\ref{R: O2 dist with C (form CO)}]-[\ref{R: O2 dist with crp}] for O$_2$. 
At high $Z'$, the H$_2$O and O$_2$ removal rates are weakly dependent on $Z'$
as the relative fractions of C, C$^+$, Si, and Si$^+$ vary. 
The overall abundances of these atoms and ions increase with
$\zeta/n$ and therefore so do the H$_2$O and O$_2$ removal rates.
The H$_2$O formation rate is independent of $Z'$
to first order, as for OH. For O$_2$ the formation rate is proportional to the product of the OH and O densities, and this is also independent of $Z'$. However, the formation rates 
increase with $\zeta/n$. Thus, in the high $Z'$ regime we expect that 
$x_{\rm H_2O}$ and $x_{\rm O_2}$ will be insensitive to 
both $Z'$ and $\zeta/n$. As seen in Figure \ref{fig: x_vs_Z}, $x_{\rm H_2O}$ and $x_{\rm O_2}$ are indeed 
roughly constant from $Z'=1$ down to 0.05. We show results as
functions of $\zeta/n$ in Figures \ref{fig: x_vs_zetan} and \ref{fig: x_vs_zetan_Z1}
below.

At lower $Z'$, and into
the atomic regime, the H$_2$O and O$_2$ 
are both removed by H$^+$. In the low $Z'$ atomic limit,
\begin{equation}
\label{eq: H2O lowZ Z and zetan dependence}
x_{\rm H_2O} \, \simeq \, 2 \times 10^{-3} \, A_{\rm O} Z' \, \left(\frac{\zeta_{-16}}{n_3} \right)^{-1}  \; \; \; ,
\end{equation} and
\begin{equation}
\label{eq: O2 lowZ Z and zetan dependence}
x_{\rm O_2} \, \simeq \, 3 \times 10^{-3} \, A_{\rm O} Z'^2 \, \left(\frac{\zeta_{-16}}{n_3} \right)^{-1.5}  \; \; \; .
\end{equation}
The H$_2$O fraction depends on the ionization parameter in the same way as OH does, and for identical reasons. For \Ot, the dependence is steeper, with one power of \z entering due to the formation via OH, and an additional $(\z)^{1/2}$ term for the proton fraction that controls the removal rate.

In the low $Z'$ limit $x_{\rm OH}$ and $x_{\rm H_2O}$ are proportional to $Z'$ because both OH and H$_2$O contain just one heavy element. However, $x_{\rm O_2} \propto Z'^2$ because O$_2$ consists of two heavy elements. These simple metallicity scalings
are inevitable when the heavy molecules are removed only by 
hydrogen-helium species and once the hydrogen-helium chemistry and the electron fraction are no longer affected by metals (or dust), as occurs at sufficiently low $Z'$ for any \z (or for sufficiently high \z for any $Z'$). In this limit, and for {\it any} network of two-body reactions, 
the abundances of any {\it trace} species containing $n$ heavy elements must always vary as $Z'^n$. Our numerical calculations are consistent with this general principle. Thus O$_2$ vanishes compared to OH in the low $Z'$ limit, as does CO compared to OH as we discuss further below.

In the molecular regime the H$_2$O/OH abundance ratio is
variable and can become large.
In the atomic regime, H$_2$O/OH is always small, $\sim 0.1$, 
(and constant) because the H$_2$O removal rate by H$^+$ is larger than for OH, and because most H$_3$O$^+$ ions dissociatively recombine to OH rather than to H$_2$O (e.g., \citealt{Vejby-Christensen1997}).


Figure \ref{fig: x_vs_Z} shows that OH/CH is a strongly decreasing function of $Z'$, even though the total oxygen to carbon abundance ratio, O$_{\rm tot}$/C$_{\rm tot}$, is independent of $Z'$ in our models. In particular, CH vanishes with decreasing $Z'$. This occurs because for all $Z' \lesssim 1$, and for our assumed $\zeta_{-16}/n_3$=1, CH is always removed by H atoms (reaction [\ref{R: CH dist}]) rather than by metal species or protons.
To a good approximation
\begin{equation}
\label{eq: CH formation}
x_{\rm CH} \, = \, \frac{k_{\rm \ref{R: CH+ form}} x_{\rm C} x_{\rm H_3^+} + k_{\rm \ref{R: CH2+ form from C+ H2}} x_{\rm C^+} x_{\rm H_2}}{k_{\rm \ref{R: CH dist}} x_{\rm H} } \; \; \; .
\end{equation} The numerator in this expression accounts for CH production initiated via the formation of CH$_2^+$ by either radiative association (reaction [\ref{R: CH2+ form from C+ H2}]) or proton transfer (reactions [\ref{R: CH+ form}]-[\ref{R: CH2+ form from CH+ H2}]) involving C or C$^+$. The denominator accounts for removal by hydrogen atoms. As $Z'$ is reduced and $x_{\rm H}$ increases, CH is rapidly removed, as seen in Figure \ref{fig: x_vs_Z}. In general CH is a negligible hydride compared to OH or H$_2$O at low metallicity.

Finally, in the middle- and lower-right panels of Figure \ref{fig: x_vs_Z} we illustrate the effects of turning on a background FUV radiation field assuming $I_{\rm UV}^0/n_3=1$.
We solve Equations (\ref{eq: chemical equilibrium x IUV on}), and as $Z'$ is reduced we vary the field intensity inside the parcels according to our dust-shielding formula  Equation (\ref{eq: IUV IUV0}). Because we are assuming that the LW band is blocked for any $Z'$, the H and H$_2$ curves are unaffected when the FUV field is turned on. However, for $Z' \lesssim 0.1$ the OH, H$_2$O, and O$_2$ are removed by photodissociation as the dust opacity vanishes. CO is fully shielded by the blocked LW band and continues to be removed by He$^+$ throughout. Nevertheless, the CO abundance is reduced when the FUV is turned on because of the suppression of the OH intermediary. Importantly, because the CO removal rate is not altered, the OH/CO abundance {\it ratio} is unaffected when the FUV is turned on, and the transition from the CO-dominated to the OH-dominated regimes still occurs near the H-to-H$_2$ transition. We discuss OH/CO further in \S\ref{sub: C/CO, OH/O2, OH/CO (illustrative cases)}.


\subsection{Dependence on $\zeta/n$ for fixed $Z'$}
\label{sub: zetan dependence}
\begin{figure}
 \includegraphics[width=.5\textwidth]{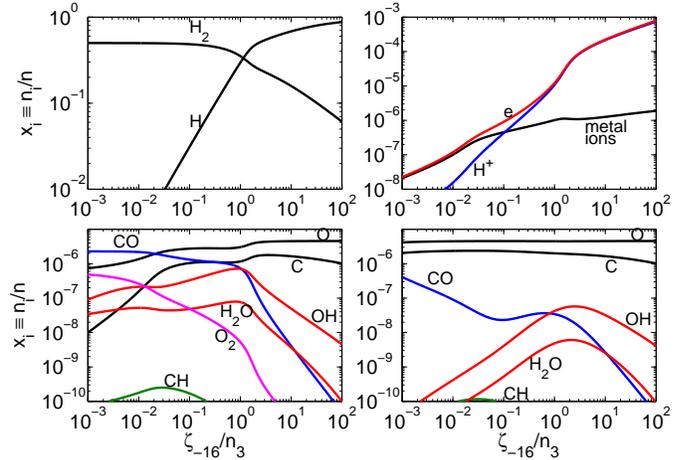}
\caption{Fractional abundances ($x_i \equiv n_{\rm{i}}/n$) as a function of $\zeta/n$ for $Z'=10^{-2}$ and $T_2=1$. In the lower right panel $I_{\rm UV}^0/n_3$=1, in all other panels $I_{\rm UV}^0/n_3$=0.}
 \label{fig: x_vs_zetan}
 \end{figure}

\begin{figure}
 \includegraphics[width=.5\textwidth]{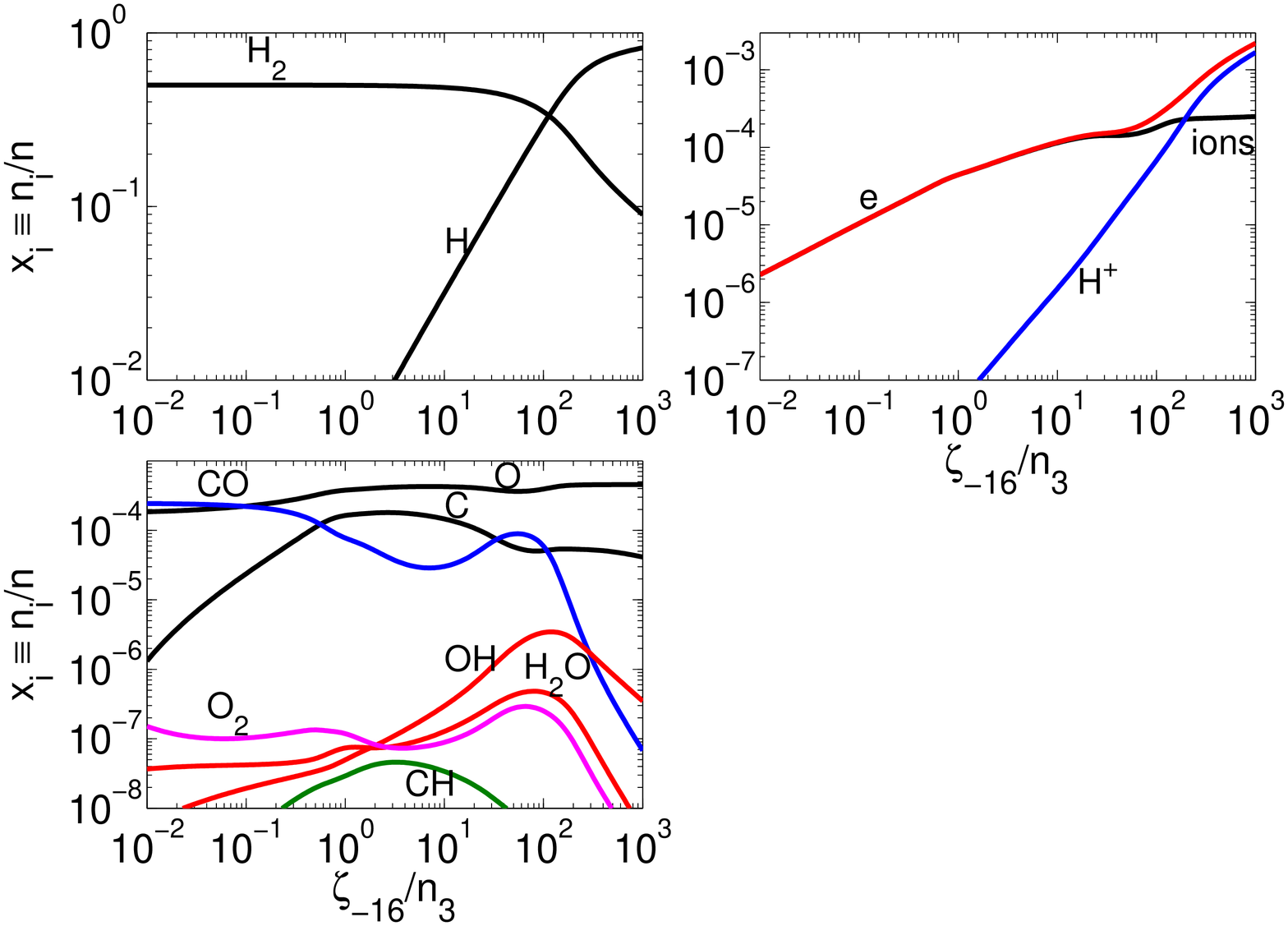}
\caption{Fractional abundances  ($x_i \equiv n_{\rm{i}}/n$) as a function of $\zeta/n$ for $Z'=1$, $T_2=1$, and $I_{\rm UV}^0/n_3$=0.}
 \label{fig: x_vs_zetan_Z1}
 \end{figure}

We now consider the behavior as a function of $\zeta_{-16}/n_3$, for two values of $Z'$. In Figure \ref{fig: x_vs_zetan}, $Z'=10^{-2}$ and \zz ranges from 10$^{-3}$ to 10$^2$. In Figure \ref{fig: x_vs_zetan_Z1}, $Z'=1$ and \zz ranges from 10$^{-2}$ to 10$^3$. Our results for $Z'=1$ are similar to those found by \citet{Lepp1996a} in their study of X-ray driven chemistry at solar metallicity, where $\zeta$ is the X-ray ionization rate (see their Figures 1 and 2). 
See also \citet{Bayet2011}, their Figure 3.

As seen in Figures \ref{fig: x_vs_zetan} and \ref{fig: x_vs_zetan_Z1}, when $\zeta/n$ is small H$_2$ destruction by crx-ionization is ineffective and the hydrogen is fully molecular (upper-left panels of Figures \ref{fig: x_vs_zetan} and \ref{fig: x_vs_zetan_Z1}). The H/H$_2$ ratio increases with $\zeta/n$, and for $Z'=10^{-2}$, the H-to-H$_2$ transition occurs at $\zeta_{-16}/n_3 \approx 1$. For $Z' = 1$ it occurs at $\zeta_{-16}/n_3 \approx 10^2$. This behavior is consistent with our Equation (\ref{eq: transition line dust}) for the location of the H-to-H$_2$ transition (for $\beta$=1).

The ionization fraction \e is shown in the upper-right panels. It also increases with $\zeta/n$. For example, for $Z'=1$, $x_{\rm e}$ ranges from $2\times10^{-6}$ to $2\times10^{-4}$ for $\zeta_{-16}/n_3$ ranging from $10^{-2}$ to $10^2$ (see also \citealt{Lepp1996a}, their Figure 1). At low $\zeta/n$, and in the H$_2$ regime, the positive charge is carried by metals. This includes molecular ions, e.g., H$_3^+$, HCO$^+$, H$_3$O$^+$, in addition to the atomic ions. At high $\zeta/n$, and in the atomic regime, the positive charge is carried by protons, and $x_{\rm e} \simeq x_{\rm H^+}$.

The lower-left panels in Figures \ref{fig: x_vs_zetan} and 
\ref{fig: x_vs_zetan_Z1} show the C, O, OH, H$_2$O, O$_2$, CH, and CO  abundance fractions $x_i$. 
The steepening rise of $x_{\rm OH}$ toward the H-to-H$_2$ transition point and then the decline with $\zeta/n$, is consistent with our approximate OH abundance scaling Equations (\ref{eq: OH highZ}) and (\ref{eq: OH lowZ Z and zetan dependence}). 
For example, for $Z'$=1, and for $\zeta_{-16}/n_3$ between $\sim 1$ and 10$^2$, $x_{\rm OH}$ increases linearly with $\zeta/n$ as given by Equation (\ref{eq: OH highZ}) for small $x_{\rm O}/x_{\rm e}$. For $\zz \lesssim 1$, the oxygen to electron ratio becomes large and the variation of $x_{\rm OH}$ with $\zeta/n$ is moderated. 
For $\zeta_{-16}/n_3 \gtrsim 10^2$ the hydrogen becomes atomic and $x_{\rm OH}$ then decreases with the ionization rate and varies approximately as $(\zeta/n)^{-1}$ as given by Equation (\ref{eq: OH lowZ Z and zetan dependence}).

Similar behavior was found by \citet{Bayet2011} in their study of molecular
chemistry at high cosmic-ray ionization rates (for models with fixed gas
density). Their numerical results also show an OH abundance peak close
to the H-to-H$_2$ transition point, but they did not discuss this connection.

At low $\zeta/n$, and in the molecular regime, O$_2$ becomes abundant compared to OH, but as $\zeta/n$ increases the O$_2$/OH ratio decreases. Correspondingly, at low $\zeta/n$ in the H$_2$ regime, most of gas-phase carbon is locked in CO, and the C/CO ratio is small. The free atomic carbon increases with $\zeta/n$ and C/CO reaches a local maximum just before the H-to-H$_2$ transition point, and the subsequent sharp drop in the CO abundance. The slight rise in the CO abundance near the H-to-H$_2$ transition in Figure \ref{fig: x_vs_zetan_Z1}, is due to the OH maximum and the resulting enhanced CO production rate at this point.

The lower right panel of Figure \ref{fig: x_vs_zetan} shows the effect of FUV radiation, for our assumed $Z'=0.01$ models for $I_{\rm UV}^0/n_3=1$. 
Photodissociation suppresses the OH, H$_2$O and O$_2$. 
At high $\zeta/n$, and well inside the atomic regime, H$^+$ dominates the OH removal and the abundance curve is unaffected by photodissociation. 
Similarly H$_2$O and CH are suppressed by photodissociation when $\zeta/n$ is not too large. CO is shielded but is indirectly reduced by the photodissociation of the OH intermediary.

For $Z'=1$ (Figure \ref{fig: x_vs_zetan_Z1}), dust-shielding is operative for all wavelengths, so that the photoprocesses are ineffective for all $\zeta/n$, and we do not include an FUV panel for this model sequence.

\subsection{Abundance Ratios: OH/O$_2$, C/CO and OH/CO}
\label{sub: C/CO, OH/O2, OH/CO (illustrative cases)}

\begin{figure}
 \includegraphics[width=.5\textwidth]{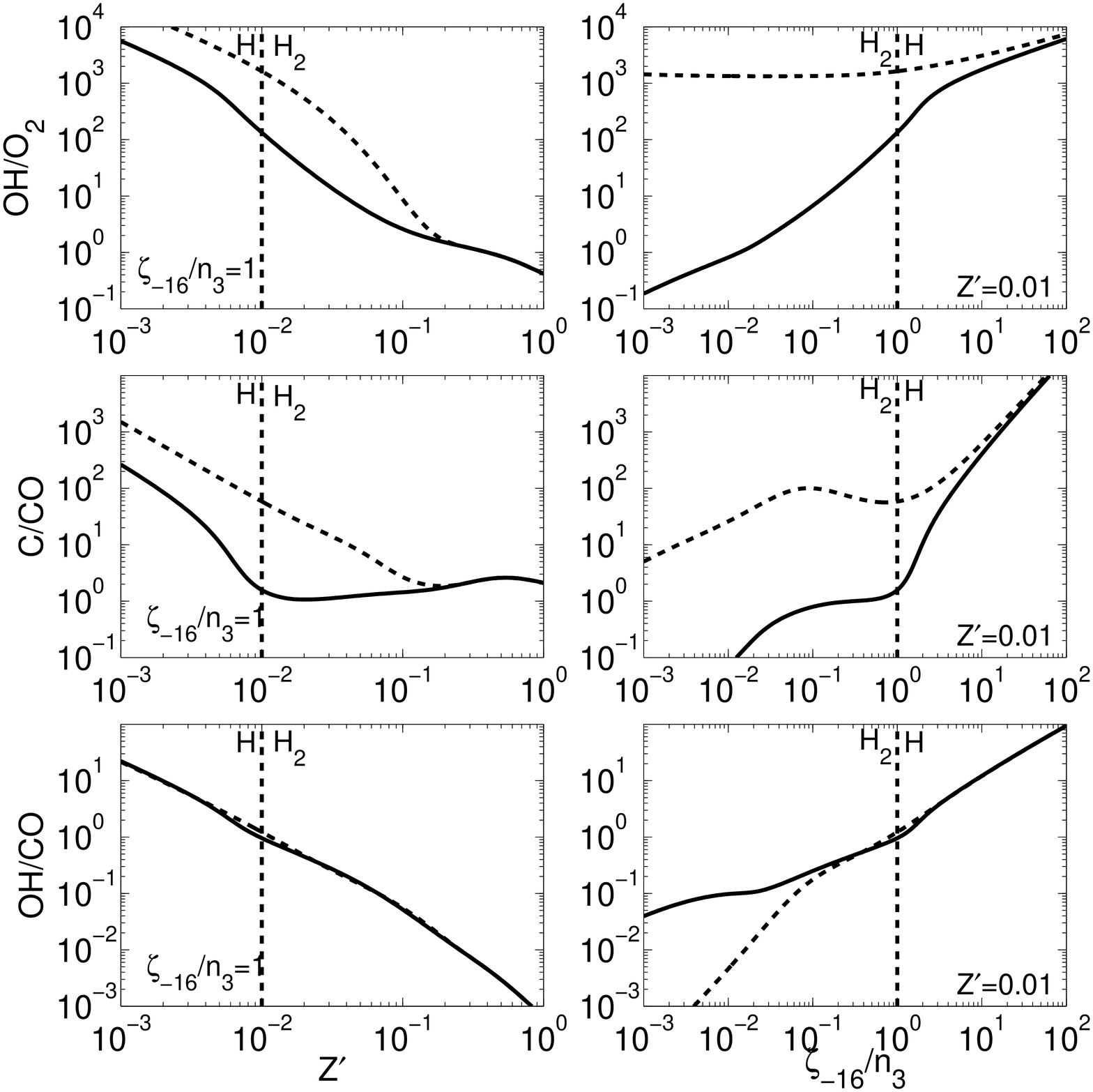}
\caption{OH/O$_2$, C/CO, and OH/CO abundance ratios as a function of $Z'$ for fixed $\zeta_{-16}/n_{3}=1$ (left) and as a function of $\zeta/n$ for fixed $Z'=10^{-2}$ (right). In both panels we assume $T_2=1$ and $I_{\rm UV}^0/n_3=0$ (solid line) and $I_{\rm UV}^0/n_3=1$ (dashed line).}
 \label{fig: OH_CO_1d}
 \end{figure}

 The C/CO, OH/O$_2$, and OH/CO abundance ratios are of particular interest. We plot them in Figure \ref{fig: OH_CO_1d}, as functions of $Z'$ for $\zeta_{-16}/n_3=1$, and as functions of \zz for $Z'=10^{-2}$, given our results presented in Figures \ref{fig: x_vs_Z} - \ref{fig: x_vs_zetan_Z1}.
 The solid curves are for FUV off, and the dashed cueves are for FUV on with $I_{\rm UV}/n_3=1$. The vertical dashed lines in Figure \ref{fig: OH_CO_1d} indicate the positions of the H-to-H$_2$ transitions. For fixed $\zeta_{-16}/n_3=1$ the H-to-H$_2$ transition occurs at $Z' \approx 10^{-2}$, and the gas is atomic (molecular) to the left (right) of the dashed line. For fixed $Z'=0.01$, the transition occurs at $\zeta_{-16}/n_3 \approx 1$, and the gas is atomic (molecular) to the right (left) of the dashed line. 
 We consider the behavior of the C/CO, OH/O$_2$, and OH/CO abundance ratios in 
 both the molecular and atomic regimes.
 
We begin with a discussion of C/CO and OH/O$_2$. These ratios are coupled in an important way in the molecular regime 
\citep{LeBourlot1993, Boger2006a, Wakelam2006}. 
To see this we first write
\begin{equation}
\label{eq: OH/O2}
\frac{\OH}{\Ot} \, \propto \, \left( \frac{\zeta_{-16}}{n_3} \right) \, \frac{1}{Z'} \; \; \; .
\end{equation}
This follows because the O$_2$ formation rate is proportional to the atomic oxygen abundance which is proportional to $Z'$, and because the O$_2$ removal rate (via reactions [\ref{R: O2 dist with C (form CO)}]-[\ref{R: O2 dist with crp}]) is approximately proportional $\zeta/n$. 
Alternatively, this is just the first-order scaling for 
$x_{\rm OH}$ with $x_{\rm O_2}$ approximately constant.
Equation (\ref{eq: OH/O2}) accounts for the fact that OH/O$_2$ can be large or  small depending on $Z'$ and $\zeta_{-16}/n_3$ as seen in Figure \ref{fig: OH_CO_1d} (and Figures \ref{fig: x_vs_Z} - \ref{fig: x_vs_zetan_Z1}).

As discussed in \S\ref{sub: C} there are two primary CO formation pathways, involving either OH as the main intermediary, via reactions [\ref{R: CO form from C,OH}]-[\ref{R: CO+ form}] with C and C$^+$, or involving O$_2$ via [\ref{R: O2 dist with C (form CO)}]-[\ref{R: O2 dist with C+ (form CO+)}], again with C and C$^+$.
In general ${\rm C^+/C \ll 1}$ in the molecular regime, 
but CO formation via C$^+$ can still contribute due to the large rate coefficients of the ion-molecule reactions. The C/CO ratio can now be be understood qualitatively by considering a simplified formation-destruction equation for the CO,
\begin{align}
\label{eq: CO simpl form dest eq}
(k_{\rm OH} \OH + k_{\rm O_2} \Ot) \C &= k_{\rm \ref{R: He+ dist High Z} } x_{\rm He^+} \CO  \nonumber\\
&= 0.5 \left( \frac{\zeta_{-16}}{n_3} \right) x_{\rm He} \, .
\end{align}
The left-hand side of this expression encapsulates CO formation, where $k_{\rm OH} \equiv 10^{-9}$~cm$^3$~s$^{-1}$ is the effective total rate coefficient for formation via OH (involving both C and C$^+$), and $k_{\rm O_2} \equiv 3 \times 10^{-10}$~cm$^3$~s$^{-1}$ is the effective rate coefficient for formation via O$_2$. We assume that C$^+$/C is small so that the abundance of carbon species not locked in CO is represented by just the free atomic carbon abundance $x_{\rm C}$. The middle term of Equation (\ref{eq: CO simpl form dest eq}) is the CO destruction rate via dissociative charge transfer with He$^+$ (reaction [\ref{R: He+ dist High Z}]). So long as this is also the dominant He$^+$ neutralization process, the rate equals the helium ionization rate on the right hand side. Because the helium is predominantly neutral we set $x_{\rm He} = A_{\rm He} = 0.1$.

At high $\zeta/n$ and/or low $Z'$ (but still in the molecular regime), OH/O$_2$ becomes large, according to Equation (\ref{eq: OH/O2}), and we then have
\begin{equation}
\label{eq: C high Z with OH}
\C \, \simeq \, \frac{0.5}{k_{\rm OH} \OH} \, \left( \frac{\zeta}{n} \right)  A_{\rm He}  \, \simeq \, 0.4 \, A_{\rm C} \, Z'  \; \; \; .
\end{equation}
Here we use our first order expression for the OH abundance in the molecular regime
\begin{equation}
\label{eq: OH 1st order mol. regime}
\OH \, \simeq \, 4 \times 10^{-8} \, \left( \frac{\zeta_{-16}}{n_3} \right) \, \frac{1}{Z'}  \; \; \; .
\end{equation}
The prefactor of $4 \times 10^{-8}$ is obtained from our numerical calculations for $\zz=1$ and $Z'=1$ (see Figure 6). Importantly, Equation (\ref{eq: C high Z with OH}) shows that when OH/O$_2 \gg 1$, \C is independent of the ionization parameter $\zeta/n$. Furthermore, \C 
approaches saturation and becomes comparable to the total carbon abundance $A_{\rm C} Z'$. In this limit the analysis shows that C/CO becomes large, but numerically we find that it remains of order unity in the H$_2$ regime.

When OH/O$_2 \ll 1$, as occurs at low $\zeta/n$ and/or high $Z'$ (again in
the molecular regime) we have
\begin{equation}
\label{eq: C high Z, with O2}
\C \, = \, \frac{0.5}{k_{\rm O_2} \Ot} \, \left( \frac{\zeta}{n} \right)  A_{\rm He} \, \simeq \, 10^{-6} \, \left( \frac{\zeta_{-16}}{n_3} \right) \ll A_{\rm C} Z' \; \; \; .
\end{equation}
Here we have assumed $\Ot \simeq 10^{-7}$, and to first-order independent of $Z'$
 and $\zeta/n$ (as found for the HIP in our full 2D computations in \S\ref{sec: parameter space}). In this limit the free atomic carbon abundance is proportional to $\zeta/n$, but C/CO remains small.

The above discussion for OH/O$_2$ and C/CO is for the H$_2$ regime. In the atomic regime, \OH and \Ot are given by Equations (\ref{eq: OH lowZ Z and zetan dependence}) and (\ref{eq: O2 lowZ Z and zetan dependence}), and
\begin{equation}
\label{eq: OH/O2 low Z}
\frac{\OH}{\Ot} \, \simeq \, 6.7 \, \left( \frac{\zeta_{-16}}{n_3} \right)^{1/2} \, \frac{1}{Z'} \; \; \; .
\end{equation} As $Z'$ becomes small OH/O$_2 \gg 1$, as expected.

In the atomic regime, the CO is formed mainly via the C+OH channel, and continues to be removed by He$^+$. Crucially however, the He$^+$ is no longer removed by CO but by charge transfer with atomic and molecular hydrogen, so that
\begin{equation}
\label{eq: He+ low Z}
x_{\rm He^+}  \, = \, \frac{0.5}{(k_{\rm \ref{R: He dist H2 to He H2+}}  + k_{\rm \ref{R: He dist H2 to He H+ H}}) \, x_{\rm H_2} +  k_{\rm \ref{R: He+ dist H}} \, x_{\rm H}  } \, \left( \frac{\zeta}{n} \right) A_{\rm He}  \; \; \; ,
\end{equation} independent of $Z'$.
In this limit
\begin{equation}
\label{eq: C/CO low Z}
\frac{\C }{\CO } \, = \, \frac{k_{\rm \ref{R: He+ dist High Z}} \, x_{\rm He^+}}{k_{\rm \ref{R: CO form from C,OH} } \, \OH } \, \simeq \, 0.5 \, \left( \frac{\zeta_{-16}}{n_3} \right)^{1.5} \, \frac{1}{Z'} \; \; \; ,
\end{equation}
and C/CO grows without limit. 
As indicated by Equations~(\ref{eq: OH/O2 low Z}) and (\ref{eq: C/CO low Z}),
in the atomic regime the OH/O$_2$ and C/CO ratios are decoupled.

We now consider OH/CO. Well inside the molecular regime where a large fraction of the carbon is locked in CO, the OH/CO ratio is small. In the atomic regime OH/CO $\gg 1$. In general the transition from the ``OH-dominated" to ``CO-dominated" regimes occurs
close to the H-to-H$_2$ transition point. This is seen in Figure \ref{fig: OH_CO_1d}.
We can write down approximate expressions for $x_{\rm OH}/x_{\rm CO}$ as follows.
At high $Z'$, and assuming most of the carbon is locked in CO, $x_{\rm CO} \propto Z'$ so that
\begin{equation}
\label{eq: OH/CO high Z'}
\frac{x_{\rm OH}}{x_{\rm CO}} \, \propto \, \frac{1}{Z'} x_{\rm OH} \, \propto 
 \, \frac{\zeta}{n} \,  \left[ 1 \, + 0.27 \, \frac{x_{\rm O}}{x_{\rm e} } \right] \, \frac{1}{Z'^2} \; \; \; ,
\end{equation} where here we have used Equation (\ref{eq: OH highZ})
for \OH. In this limit OH/CO varies as 1/$Z'^2$, and sublinearly with $\zeta/n$.

In the atomic regime CO is formed via the OH+C channel, and is still
removed by He$^+$, so that
\begin{equation}
\label{eq: OH/CO low Z' no proportionality relations}
\frac{x_{\rm OH}}{x_{\rm CO}} \, = \, \frac{k_{\rm \ref{R: He+ dist High Z}} x_{\rm He^+}}{k_{\rm \ref{R: CO form from C,OH}} x_{\rm C}} \, \simeq 10^{-2} \left( \frac{\zeta_{-16}}{n_3} \right) \, \frac{1}{Z'} \; \; \; ,
\end{equation}
where $\C = A_{\rm C} Z'$, and $x_{\rm He^+}$ is given by Equation (\ref{eq: He+ low Z}). Once again, because
OH contains one heavy element and CO consists of two, OH/CO varies as 1/$Z'$ in the low $Z'$ limit. The linear dependence on \z enters via $x_{\rm He^+}$
as given by Equation~(\ref{eq: He+ low Z}).

The dashed curves in Figure 9 are for FUV on with $I_{\rm UV}^0/n_3=1$. The OH/O$_2$ ratio is now inversely proportional to the O$_2$ photodestruction rate and is independent of $\zeta/n$, but the dependence on $Z'$ is unaffected. C/CO is enhanced due to the indirect reduction of CO via photodissociation of the OH and the O$_2$. 
Importantly, so
long as CO is mainly formed via the OH intermediary, OH/CO is unaffected by photodissociation of the OH. This is seen in the lower panels of Figure \ref{fig: OH_CO_1d}, in which the dashed (FUV on) and solid (FUV off) curves coincide. The OH/CO curves do diverge for very low $\zeta/n$ at fixed $Z'$, because when FUV is on the OH (and O$_2$) abundances become so small that residual processes start contributing to the CO production. However in the (atomic) ``OH-dominated" regime photodissociation of the OH never affects the OH/CO ratio. In particular, 
photodissociation does not affect the transition points from the CO-dominated to OH-dominated regimes.

\section{Chemical Abundances for $Z'$ Versus \z}
\label{sec: parameter space}
Given our description in \S\ref{sec: H_H2_analytical} and \ref{sec: Two illustrative} of the basic chemical behavior and regimes, we now present computations that span our full two-dimensional (2D) $Z'$ versus $\zeta/n$ parameter space. For our 2D computations we vary $Z'$ from 10$^{-3}$ to 1 and $\zeta_{-16}/n_3$ from 10$^{-3}$ to 10$^2$. 
We again assume that the grain H$_2$ formation rate coefficient varies linearly with $Z'$ ($\beta=1$ in Equation [\ref{eq: grain rate}]), and we show results for FUV off (\S\ref{sub:2d H, H2, Hp, e} and \ref{sub: 2d C, CO, OH, H2O, O2}) and on (\S\ref{sub: FUV on 2d}), according to Equations (\ref{eq: mass conservation})-(\ref{eq: chemical equilibrium x}), and (\ref{eq: chemical equilibrium x IUV on})-(\ref{eq: IUV IUV0}). We again assume $T_2=1$ gas.
We present the atomic and molecular abundances in the $Z'$ versus \zz plane as color-contour plots, in which the color varies with $\log \, x_i$, from large (red) to small (blue). The displayed dynamic-ranges for the $\log \, x_i$ differ for different species $i$, and are indicated for each panel.


\begin{figure}
 \includegraphics[width=.5\textwidth]{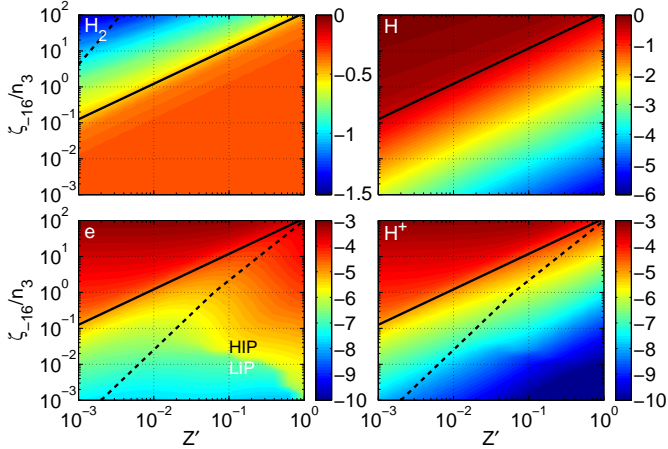}
\caption{Fractional abundances in the $Z'$ versus $\zeta_{-16}/n_3$ plane, for $I_{\rm UV}=0$ and $T_2=1$. (a) Upper left panel shows $\log \, x_{\rm H_2}$, ranging from -1.5 to 0. The solid line is the H-to-H$_2$ transition where $x_{\rm H_2}=x_{\rm H}$ (shown in all panels), and the dashed line is where gas phase and dust grain H$_2$ formation rates equal. (b) Upper right panel is $\log \, x_{\rm H}$, from -6 to 0. (c) Lower left panel is $\log \, x_{\rm e}$, from -10 to -3. The abrupt switch from HIP to LIP is indicated. (d) Lower right panel is $\log \, x_{\rm H^+}$, from -10 to -3, the dark blue area indicates values lower than the color-bar range. In both lower panels, the dashed curves are where $x_{\rm H^+} = 0.25 x_{\rm e}$.}
 \label{fig: 2d_H2}
 \end{figure}
 
\subsection{H, H$_2$, H$^+$, and e}
\label{sub:2d H, H2, Hp, e}
In Figure \ref{fig: 2d_H2} we show the atomic and molecular hydrogen fractions \H{} and \H{2}, and the electron and proton fractions, \e and \Hp{}, in the $Z'$ versus $\zeta_{-16}/n_3$ plane assuming FUV off ($I_{\rm UV}^0=0$), as given by the solutions to Equations (\ref{eq: mass conservation})-(\ref{eq: chemical equilibrium x}).

The upper left panel shows \H{2} for the range $10^{-1.5}$ to 1. The upper right panel shows \H{} from $10^{-6}$ to 1. The solid diagonal line shows the H-to-H$_2$ transition where \H{}=\H{2}. Its slope, $(\zeta/n)\propto Z'$, and position is consistent with our Equation (\ref{eq: transition line dust}).
We draw the H-to-H$_2$ transition line in all of our parameter-space plots to delineate the atomic and molecular hydrogen regimes. As seen in Figure \ref{fig: 2d_H2}, the gas becomes atomic at a given $Z'$ for sufficiently large $\zeta/n$, or at sufficiently low $Z'$ for a given $\zeta/n$. In the panel for $x_{\rm H_2}$, we also draw a dashed line where the H$_2$ formation rates in the gas-phase and on grains are equal. The position of this line is in good agreement with our Equation (\ref{eq: transition between gas and dust}) for $\beta=1$. Grain H$_2$ formation is dominant in most of our parameter space including in the atomic regime, except in the upper-left corner where $Z'$ is very low and $\zeta_{-16}/n_3$ is large. 

The lower  left and right panels of Figure \ref{fig: 2d_H2} shows \e and H$^+$ ranging from 10$^{-10}$ to 10$^{-3}$. Several zones may be identified for the fractional ionization $x_{\rm e}$. The solid line again divides the parameter-space into the atomic and molecular regimes. 
The abrupt switch from LIP to HIP conditions can be seen
at $\zeta_{-16}/n_3 \approx 10^{-2}$ to 10$^{-3}$. The dashed curve is where 
$x_{\rm H^+} = 0.25 x_{\rm e}$. To the left of this curve H$^+$ becomes the dominant positive charge carrier, and in the atomic regime
$x_{\rm e}$ varies as $(\zeta_{-16}/n_3)^{1/2}$
(see Equation [\ref{eq: xe}]). 
To the right of the dashed curve metals dominate the fractional ionization. Thus,
for a given $\zeta_{-16}/n_3$, \e is maximal at very low $Z'$, and decreases across the H-to-H$_2$ transition as $x_{\rm H^+}$ drops. The electron fraction then increases again across the dashed curve, as the metals take over.
This is the behavior seen for \e versus $Z'$ in our fiducial 1D cut shown in Figure \ref{fig: x_vs_Z} for $\zeta_{-16}/n_3=1$. At fixed $Z'$, $x_{\rm e}$ increases monotonically with $\zeta/n$, as we have also seen in Figures \ref{fig: x_vs_zetan} and \ref{fig: x_vs_zetan_Z1}.

    \begin{figure}
 \includegraphics[width=.5\textwidth]{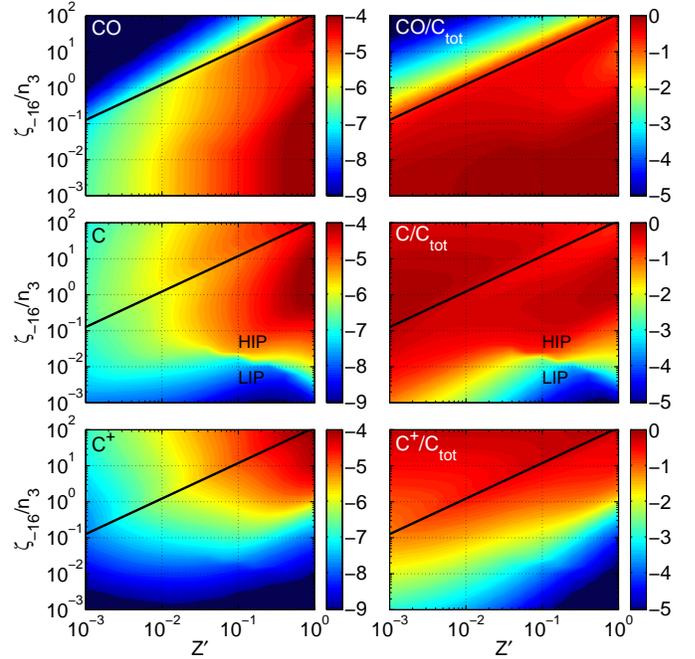}
\caption{CO, C, and C$^+$ in the $Z'$, $\zeta_{-16}/n_3$ plane with $T_2=1$ and $I_{\rm UV}^0=0$. The left panels show logarithmic abundances relative to total hydrogen density, $\log \, x_{\rm CO}$, $\log \, x_{\rm C}$ and $\log \, x_{\rm C^+}$ ranging from -9 to -4. The right panels show the $\log$ abundances relative to the total elemental carbon (C$_{\rm tot} \equiv 2.9 \times 10^{-4} Z'$), ranging from -5 to 0. The dark blue (red) areas indicate values lower (higher) than the color-bar range. The H-to-H$_2$ transition is at the solid line. The LIP and HIP zones are indicated.}
 \label{fig: 2d_CO_C_UV0}
 \end{figure}
 
     \begin{figure}
 \includegraphics[width=.5\textwidth]{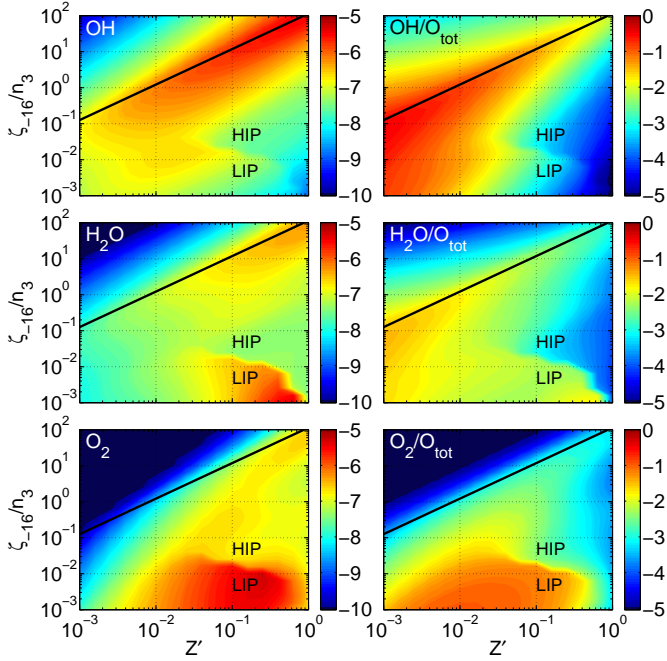}
\caption{OH, H$_2$O, and O$_2$ in the $Z'$, $\zeta_{-16}/n_3$ plane with $T_2=1$ and $I_{\rm UV}^0=0$. The left panels show logarithmic abundances relative to total hydrogen density, $\log \, x_{\rm OH}$, $\log \, x_{\rm H_2O}$ and $\log \, x_{\rm O_2}$, ranging from -10 to -5. The right panels show the $\log$ abundances relative to the total elemental oxygen (O$_{\rm tot} \equiv 4.9 \times 10^{-4} Z'$), ranging from -5 to 0. The dark blue (red) areas indicate values lower (higher) than the color-bar range. The H-to-H$_2$ transition is at the solid line. The LIP and HIP zones are indicated.}
 \label{fig: 2d_OH_H2O_UV0}
 \end{figure}

\subsection{CO, C, C$^+$, OH, H$_2$O and O$_2$}
\label{sub: 2d C, CO, OH, H2O, O2}
Our 2D results for the C, CO, OH, H$_2$O and O$_2$ abundances,
with FUV off, are shown in Figures \ref{fig: 2d_CO_C_UV0} and \ref{fig: 2d_OH_H2O_UV0}. In the left hand panels we show the $x_i$.
In the right hand panels we normalize to the total carbon or oxygen abundances, 
C$_{\rm tot}=A_{\rm C} Z'$ and O$_{\rm tot} = A_{\rm O} Z'$. 
In Figure \ref{fig: 2d_CO_C_UV0} we show CO/C$_{\rm tot}$, C/C$_{\rm tot}$ and C$^+$/C$_{\rm tot}$, and in Figure \ref{fig: 2d_OH_H2O_UV0} we show
OH/O$_{\rm tot}$, H$_2$O/O$_{\rm tot}$, and O$_2$/O$_{\rm tot}$.

The upper panels of Figure \ref{fig: 2d_CO_C_UV0} display the CO abundances,
with $x_{\rm CO}$ ranging from 10$^{-4}$ to $\sim 10^{-7}$ in the molecular
regime (below the solid line) and down to less than 10$^{-9}$ in the atomic regime. 
These plots show
that in the H$_2$ regime a large fraction, 30-100~\%, of the available carbon is always locked in CO. This is true even at very low $Z'$, so long as \z is small enough for the hydrogen to be molecular. The general result is that gas-phase CO formation is efficient at all metallicities so long as the hydrogen is molecular.
Thus, in the H$_2$ regime $x_{\rm CO}$ varies approximately linearly with $Z'$ and very weakly with $\zeta/n$. However, in the atomic regime \CO drops very sharply with decreasing $Z'$ or increasing $\zeta/n$, and CO becomes a trace species. In this regime $\CO \propto (\z)^{-1.5} Z'^2$ as seen in Figure \ref{fig: 2d_CO_C_UV0} and consistent with Equation (\ref{eq: C/CO low Z}).

The middle and lower panels of Figure \ref{fig: 2d_CO_C_UV0} show the free atomic and ionized carbon abundances. Even in the H$_2$ regime, a significant fraction ($\sim 50\%$) of the carbon remains atomic as long as $\zeta/n$ is not too low, especially in the HIP. 
But  at sufficiently low $\zeta/n$ for any $Z'$ the atomic carbon vanishes and is fully absorbed into CO. This is the behavior we discussed in \S\ref{sub: C/CO, OH/O2, OH/CO (illustrative cases)}. In the atomic regime where CO vanishes most of the carbon remains atomic as long as the ionization parameter is not too large, with C$^+$/C$<1$ for $\zz \lesssim 50$. For higher ionization parameters most of the carbon is ionized. 

The upper panels of
Figure \ref{fig: 2d_OH_H2O_UV0} show the behavior for OH. 
To first order, $x_{\rm OH} \propto (\zeta/n)/Z'$ in the molecular
regime (see Equation[\ref{eq: OH highZ}]) and the color contours for $x_{\rm OH}$
run approximately parallel to the H-to-H$_2$ transition line.
As we have discussed, for any $Z'$ we expect the OH abundance to be maximal 
near the H-to-H$_2$ transition. This is indeed 
seen in Figure \ref{fig: 2d_OH_H2O_UV0}.
Along the transition line, $x_{\rm OH}$ ranges from $\approx 3\times 10^{-6}$
at $Z'=1$ to $\approx 10^{-7}$ at $Z'=10^{-3}$. 
The reduction of \OH along the transition line is due to H$^+$ removal of the OH (reaction [\ref{R: OH dist lowZ}]) in addition to removal by atomic oxygen and other metals.
Along the transition line the OH is formed mainly via the charge-transfer route with H$^+$ (reaction [\ref{R: O+ form}]). Well within the molecular regime where H$^+$ becomes small the H$_3^+$ proton transfer route predominates (reaction [\ref{R: OH+ form with H3+}]). The jump in \OH at the HIP-to-LIP boundary is due, in part, to the sudden change in the H$_3^+$ abundance at the phase transition.
In the atomic regime the OH abundance decreases, and $\OH \propto (\zeta/n)^{-1} Z'$ (see Equation [\ref{eq: OH lowZ Z and zetan dependence}]), but the relative fraction OH/O$_{\rm tot}$ is relatively large in this regime, as seen in the upper-right panel.

The middle panels of Figure \ref{fig: 2d_OH_H2O_UV0} show that $x_{\rm H_2O}$ is weakly dependent on $Z'$ and $\zeta/n$ in the H$_2$ regime. We find $x_{\rm H_2O} \approx 10^{-8}$ to $10^{-7}$ in the HIP and $\sim 3 \times 10^{-6}$ in the LIP.
In the atomic regime the H$_2$O abundance decreases, and
$x_{\rm H_2O}\propto (\zeta/n)^{-1} Z'$ (see Equation [\ref{eq: H2O lowZ Z and zetan dependence}]).

     \begin{figure}
 \includegraphics[width=.5\textwidth]{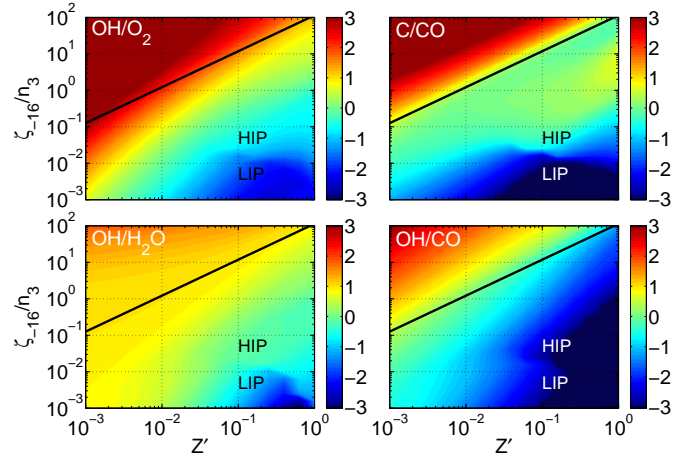}
\caption{$\log$ of the abundance ratios, OH/O$_2$, C/CO, OH/H$_2$O and OH/CO in the $Z'$, $\zeta_{-16}/n_3$ plane ranging from -3 to 3, for $T_2=1$ and $I_{\rm UV}^0=0$. The dark blue (red) areas indicate values lower (higher) than the color-bar range. The H-to-H$_2$ transition is at the solid line. The LIP and HIP zones are indicated.}
 \label{fig: 2d_ratios_UV0}
 \end{figure}

The lower panels of Figure \ref{fig: 2d_OH_H2O_UV0} show that as expected
$x_{\rm O_2}$ is also weakly dependent on $Z'$ and $\zeta/n$ in much of the
H$_2$ regime. 
In the HIP $\Ot \approx 2 \times 10^{-7}$ is a characteristic value. For the LIP $\Ot \approx 2 \times 10^{-6}$. 
   \begin{figure}
 \includegraphics[width=.25\textwidth]{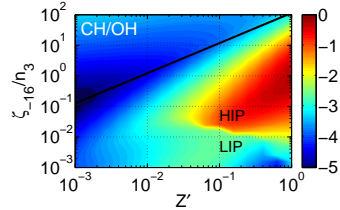}
\caption{$\log$ of the CH/OH abundance ratio in the $Z'$, $\zeta_{-16}/n_3$ plane, ranging from -5 to 0, for $T_2=1$ and $I_{\rm UV}^0=0$. The dark blue (red) areas indicate values lower (higher) than the color-bar range. The H-to-H$_2$ transition is at the solid line.}
 \label{fig: CH_OH}
 \end{figure}
 At sufficiently low $Z'$ removal
of the O$_2$ by H$^+$ rather than metals takes over, and $x_{\rm O_2}$ drops.
In the atomic regime the O$_2$ rapidly vanishes as $(\zeta/n)^{-1.5} Z'^2$
(see Equation [\ref{eq: O2 lowZ Z and zetan dependence}]).

Figure \ref{fig: 2d_ratios_UV0} shows the abundance ratios for OH/O$_2$, C/CO, OH/H$_2$O, and OH/CO, in our $Z'$ versus $\zeta/n$ parameter space. 
In these plots we limit the dynamic ranges 
of the displayed ratios from $10^{-3}$ to $10^3$. 

Most of the variation in OH/O$_2$ is due to the variation in $x_{\rm OH}$.
At low $\zeta/n$ and/or at high $Z'$, in the molecular regime, OH/O$_2$ becomes small (see Equation [\ref{eq: OH/O2}]). Still in the molecular regime, OH/O$_2$ approaches unity as $Z'$ is decreased or as $\zeta/n$ is increased. For example, 
OH/O$_2 \approx 1$ at $Z' = 3 \times 10^{-3}$ for $\zeta_{-16}/n_3=10^{-3}$ or at $Z'=1$ for $\zeta_{-16}/n_3=20$. At still lower $Z'$, or higher $\zeta/n$, OH begins to dominate over O$_2$. 
In the atomic regime OH {\it always} strongly exceeds O$_2$ as it must, due to the $Z'$ versus $Z'^2$ metallicity scalings for the OH and O$_2$ abundances in the low $Z'$ limit.

As discussed in \S\ref{sub: C/CO, OH/O2, OH/CO (illustrative cases)}, the C/CO and OH/O$_2$ abundance ratios are strongly coupled in the molecular regime. 
So long as OH/O$_2$ is not too low C/CO remains of order unity. At very low $\zeta/n$, OH/O$_2$ becomes small and consequently C/CO strongly decreases (approximately as $\zeta/n$) as all of the carbon is driven to CO. 
For example, at solar metallicity, C/CO $\approx 0.1$ for $\zeta_{-16}/n_3=10^{-1}$, but ${\rm C/CO} \approx 6 \times 10^{-3}$ for $\zeta_{-16}/n_3=10^{-2}$. In the atomic regime, C/CO rapidly increases as  the CO vanishes.

The lower-left panel of Figure \ref{fig: 2d_ratios_UV0} shows OH/H$_2$O. In the atomic regime OH/H$_2$O $\sim 10$ and is approximately constant, as set by the branching ratios of H$_3$O$^+$ dissociative recombination, and the preferential removal of H$_2$O by protons. In the molecular regime, OH/H$_2$O varies and
can become small as \OH decreases more rapidly at high $Z'$ and low $\zeta/n$. 

The OH/CO ratio for $\zeta/n$ versus $Z'$ is
shown in the lower-right panel of Figure \ref{fig: 2d_ratios_UV0}. 
As expected, OH/CO$\ll 1$ well inside the H$_2$ regime.
In the atomic regime OH/CO~$\gg 1$.
The switch from the OH- to CO-dominated regimes occurs close
to the H-to-H$_2$ transition line.

Finally in Figure \ref{fig: CH_OH} we show the CH/OH abundance ratio within our $Z'$ versus \z parameter space. The CH/OH ratio approaches unity at high $Z'$ within the HIP. As expected, CH vanishes and CH/OH becomes small at low $Z'$ and into the atomic regime. OH is the dominant heavy molecule at low $Z'$ and/or high $\zeta/n$.

\subsection{FUV on}
\label{sub: FUV on 2d}
\begin{figure}
 \includegraphics[width=.5\textwidth]{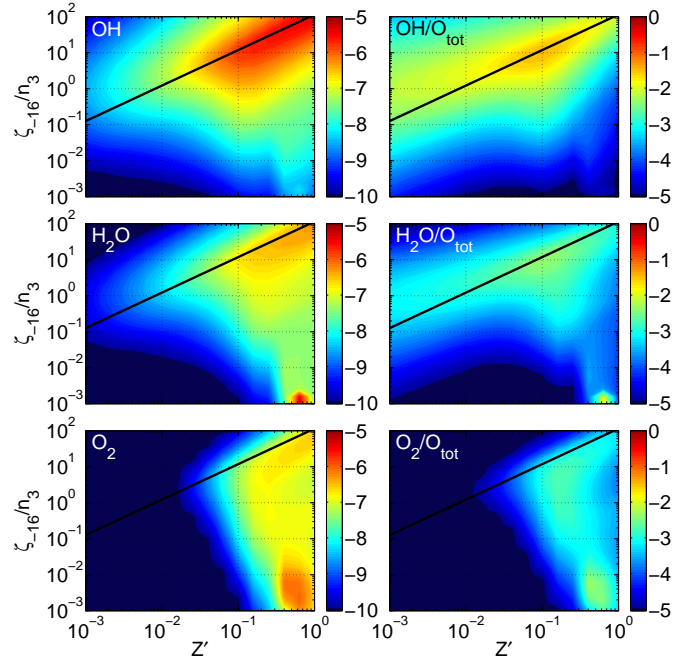}
\caption{The same as Figure \ref{fig: 2d_OH_H2O_UV0} but with $I_{\rm UV}^0/n_3=1$.}
 \label{fig: 2d_OH_H2O_UV1}
 \end{figure}

\begin{figure}
 \includegraphics[width=.5\textwidth]{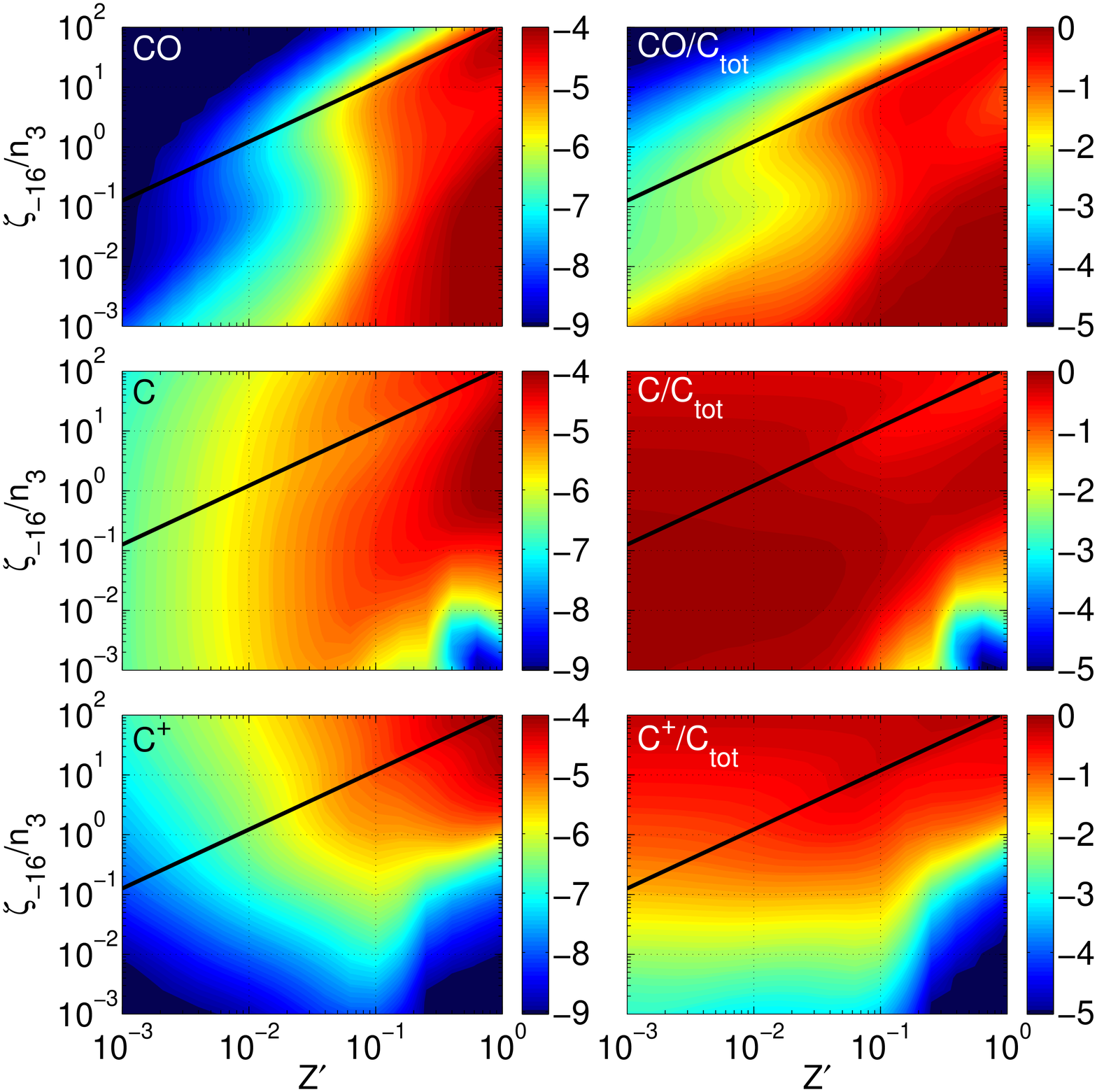}
\caption{The same as Figure \ref{fig: 2d_CO_C_UV0} but with $I_{\rm UV}^0/n_3=1$.}
 \label{fig: 2d_CO_C_UV1}
 \end{figure}

\begin{figure}
 \includegraphics[width=.5\textwidth]{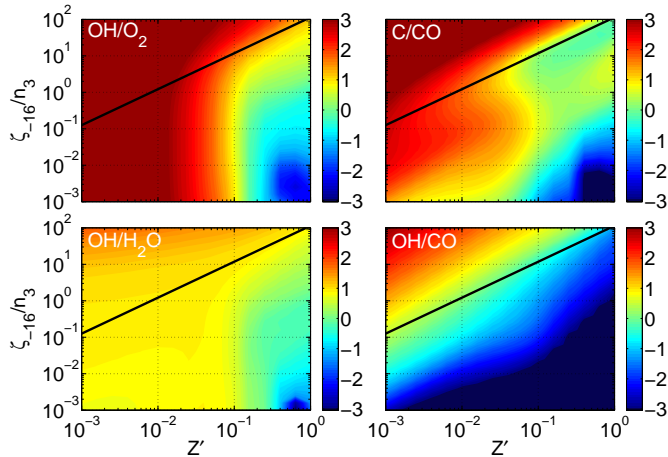}
\caption{The same as Figure \ref{fig: 2d_ratios_UV0} but with $I_{\rm UV}^0/n_3=1$.}
 \label{fig: 2d_ratios_UV1}
 \end{figure}

In Figures \ref{fig: 2d_OH_H2O_UV1}, \ref{fig: 2d_CO_C_UV1} and \ref{fig: 2d_ratios_UV1}, we show 2D model results for FUV on.
We again assume $I_{\rm UV}^0/n_3=1$,
and vary $I_{\rm UV}$ inside the parcels according to our dust shielding formula (Equation [\ref{eq: IUV IUV0}]). 
As before, we assume that the LW band is always completely blocked so that the H$_2$ and CO are shielded from direct photodissociation at all $Z'$. However OH, H$_2$O, and O$_2$ are photodissociated when $Z'$ is small and the dust shielding vanishes. 
For our assumed $I_{\rm UV}^0/n_3=1$ photodissociation dominates over chemical
removal for $Z' \lesssim 0.1$ for the HIP, and for $Z' \lesssim 0.4$ for the LIP.

Figure \ref{fig: 2d_OH_H2O_UV1} shows the OH, H$_2$O, and O$_2$ abundances with FUV on. For $\zz \gtrsim 1$, chemical removal remains dominant at all $Z'$. For lower ionization parameters, photodissociation becomes rapid once the parcels become unshielded at $Z' \lesssim 0.1$, and OH, H$_2$O, and O$_2$ abundances are reduced. 
Reduction of O$_2$ by FUV is especially effective because (a) the OH intermediary required for O$_2$ formation is removed and (b) the O$_2$ is itself directly photodissociated.

The upper panel of Figure \ref{fig: 2d_CO_C_UV1} shows that the CO vanishes within the H$_2$ regime as the OH (and O$_2$) are photodissociated, even though the CO itself is fully shielded. CO then becomes a trace species as $Z'$ is reduced, and most of the carbon remains atomic or ionized as shown in the middle panels.

Figure \ref{fig: 2d_ratios_UV1} shows the OH/O$_2$, C/CO, OH/H$_2$O, and OH/CO abundance ratios for the computations with FUV on.
The OH/O$_2$ ratio becomes very high in most of the parameter space. 
When photodissociation is effective we find OH/H$_2$O $\approx 10$, now also in the H$_2$ regime, not just in the atomic regime. 
This is now due to preferential formation of OH in the dissociative recombination of H$_3$O$^+$, and the higher H$_2$O photodissociation rate compared to OH.
Once photodissociation becomes effective the OH/O$_2$ and C/CO ratios decouple, and the C/CO ratio becomes large as the CO vanishes.

Importantly, the OH/CO ratio is unaffected by the FUV, because only the OH, that is required for the formation of the CO, is reduced by the FUV. Thus the transition from the CO to OH dominated regimes still occurs close to the H-to-H$_2$ transition line (as given by Equation [\ref{eq: transition line dust}]), whether or not FUV is off or on. 
Inevitably, OH becomes the dominant heavy molecule in the limit of low metallicity.

\section{Comparison to Time-dependent Molecule Formation Sequences} 
\label{sec:Comparison}
 Before concluding it is of interest to compare our results
 assuming equilibrium ionization-driven chemistry for the 
 the bulk ``dense-cold-ISM",
to the time-dependent calculations presented 
by \citet{Omukai2005} in their study 
of collapsing clouds at low metallicities.  
As we noted in \S 1,
in the collapsing cloud calculations
 the chemistry consists of intrinsically time-dependent reformation sequences, and is not ionization-driven as in our study.
Because ionization (and/or photodissociation) by external
driving sources is not included in the 
collapsing cloud computations, these systems can never formally
reach a formation-destruction equilibrium.
Nevertheless, some qualitative comparisons can be made. 

For this purpose we consider the Omukai et al.~abundance versus 
density curves (their Figs. 5 and 12) for H and H$_2$, for
C$^+$, C, and CO, and for O, OH, and H$_2$O.  (They do not
display results for CH and other carbon hydrides.) 
In their scheme the gas density, $n(t)$, increases with time $t$
and the non-equilibrium chemical abundances 
vary with gas density, which is a proxy for time.
Omukai et al.~display the chemical abundances as functions
of density, but the times associated with each density 
and chemical state are not shown.
The precise behavior
depends on initial conditions and the varying temperatures as set by compressive
heating and radiative cooling.  For a very wide range of 
gas densities (10$^2$ to 10$^8$ cm$^{-3}$) and metallicities
(down to $Z' \sim 10^{-3}$) the gas temperature is in our
cold gas ($T\sim 100$~K) regime, and the chemistry is
dominated by the two-body reactions.

The time-dependent curves show that the atomic to molecular
hydrogen conversion points occur at larger densities
for lower metallicities, as is expected given their
($\beta=1$) assumption that the H$_2$ grain formation
time scale varies inversely with $Z'$.  They also find
that once the hydrogen is fully molecular, the conversion
to CO is always complete.  This is qualitatively similar to
our finding that CO production is efficient in the H$_2$ regime.
However, in the reformation sequences, conversion 
to CO occurs before the H/H$_2$ transition, and CO
exists within part of the atomic regime.  This probably 
reflects the fact that He$^+$ has recombined and the CO
removal rates have become vanishingly small
(with photodissociation also excluded). For our
ionization-driven chemistry we find that  
CO always disappears close to the H/H$_2$ transition point
as set by the ionization parameter \z (as opposed to a
time-dependent density $n$). Furthermore, in our models CO
is not always fully formed even within the H$_2$ regime.

Second, in the contracting clouds OH/CO$\ll 1$ in the
H$_2$ regime.  However, it is unclear how the OH/CO ratio would
be affected by steady background ionization
\footnote{\citet{Omukai2012} include background ionization in their
revised cooling/fragmentation computations, but do not present
results for the chemistry.}, which also enables
a well-defined reference equilibrium state for any
density and temperature.  In the (early time) atomic regimes, 
OH/CO is large, although not as large
as given by our  OH/CO$\sim 1/Z'$ scaling rule.
As we have demonstrated mathematically, this rule
must hold for low-$Z'$ ionization-driven sequences
in the atomic regime.

Omukai et al.~also find that OH/H$_2$O $>$ 1 in the cold-gas
regime as is expected for OH and H$_2$O formation via dissociative recombination.  They did not present results for other hydrides,
in particular CH, which could in principle compete with OH. However,
as we have shown, at low $Z'$ into the atomic regime, 
CH vanishes compared to OH even for comparable carbon and oxygen elemental abundances.

Our study of ionization-driven molecular equilibrium states
down to low metallicity is complementary to the
time-dependent molecular formation sequences in collapsing clouds.
An analysis of low-metallicity time-dependent formation chemistry
with the inclusion of background irradiation and ionization would
be valuable, and will be presented in forthcoming papers.

\section{Summary and Discussion}
\label{sec: Discussion}


In this paper we have presented a numerical and analytic
study of interstellar gas-phase ion-molecule chemistry in cold dense gas, 
from solar metallicity Galactic
conditions into the domain of interstellar media at very low metallicities, 
and related properties at 
high ionization rates. Relevant astrophysical environments include
the cool ISM in low metallicity dwarf galaxies, early enriched clouds
at reionization and the Pop-II star formation epoch, and in dense
gas exposed to intense Xray or cosmic-ray sources.
We consider a wide range of overall heavy-element abundances or 
metallicities, $Z'$, and cloud ionization parameters $\zeta/n$.
Here $\zeta$ is the total ionization rate, and $n$ is the
hydrogen gas volume density.

We have focused on the steady-state behavior and trends for 
H$_2$, CO, CH, OH, H$_2$O, and O$_2$, and the associated 
diagnostic abundance ratios CO/H$_2$, C/CO, OH/CO, 
and OH/O$_2$. In \S \ref{sec: Network} we present a detailed discussion of the
hydrogen-carbon-oxygen chemical networks and the
varying molecule formation-destruction pathways, 
down to the low metallicity regime.

We have considered idealized shielded or partially 
shielded ``one-zone" isothermal gas parcels,
for which the steady-state abundances of the atomic and molecular 
species are fully determined by the assumed metallicity
and ionization parameter, as discussed in \S \ref{sec: Model ingredients}.
We adopt a gas temperature of 100~K appropriate for cooled gas
at low metallicity.
We imagine that the (hydrogen) ionization rate $\zeta$ is 
provided by penetrating cosmic-rays or X-rays, and the associated 
secondary electrons. When FUV photo(destruction) processes are 
included a third parameter enters, the ratio $I^0_{\rm UV}/n$ of the 
FUV intensity to the gas density. We have then considered two types
of models. First are sequences for which irradiation by
external FUV is fully excluded.
Second are models for which FUV is turned on, but the H$_2$
Lyman-Werner photodissociation band remains fully blocked 
even when the parcels are optically thin to longer wavelength radiation.
Our overall parameter space is, $10^{-3} \leqslant Z' \leqslant 1$ 
and $10^{-3} \leqslant \zeta_{-16}/n_3 \leqslant 10^{2}$.
We assume a solar relative abundance pattern
for the heavy elements, with absolute total abundances 
proportional to $Z'$.
For models with FUV-on we set $I_{\rm UV}/n_3=1$. 
We adopt relative photorates appropriate for a diluted
10$^5$~K blackbody field as representative of pop-III star spectra.
Our computed atomic and molecular photodissociation
and photoionization rates for such spectra are presented in Table 2.

The H/H$_2$ balance is critical to the overall chemical behavior
since the ion-molecule sequences for the metal-bearing 
species all require at least some H$_2$. In \S\ref{sec: H_H2_analytical} we present general 
purpose analytic results for the H-to-H$_2$ transition points and
H$_2$ fractional abundances for optically thick conditions (i.e.~no
LW photodissociation) in which the H$_2$ is destroyed 
by ionization (by cosmic-rays or X-rays) and is formed
on dust-grains or in the gas phase via the H$^-$ negative ion.
We assume that the dust-to-gas ratio and
H$_2$ grain formation efficiency vary as a simple power-law
of the metallicity. The ionization parameters at the 
H-to-H$_2$ transition points in the limits of grain versus gas-phase
H$_2$ formation are given by our Equations (\ref{eq: transition line gas}) and (\ref{eq: transition line dust}). 
The H$_2$ abundances (in the atomic regime) 
are given by Equations (\ref{eq: xH2 gas}) and (\ref{eq: xH2 dust}).
We have verified our H/H$_2$ analysis with our detailed 
full chemistry computations in \S \ref{sec: Two illustrative} and \ref{sec: parameter space}. 

For our chemical study we have first considered 1D cuts,
for varying $Z'$ and fixed $\zeta/n$, and second for varying
$\zeta/n$ at fixed $Z'$, with and without FUV. 
These computations are presented in \S \ref{sec: Two illustrative}, and enable
our development of abundance scaling formulae, 
mainly for OH, O$_2$ and CO,
in both the molecular and atomic hydrogen regimes,
from high to low metallicity.
The division between these regimes
in our full $Z'$ versus $\zeta/n$ parameter space depends
on how the dust-to-gas ratio varies with metallicity (\S\ref{sec: H_H2_analytical}).
In our numerical calculations (\S\ref{sec: Two illustrative} and \ref{sec: parameter space}) we have assumed a linear variation. 

In \S \ref{sec: parameter space} we present the results of our chemical computations 
as color-contour plots in our 2D, $Z'$ versus $\zeta/n$
parameter space. Our primary results, all for steady state
conditions, are as follows. First, in the absence of photodissociation 
CO molecules always form efficiently even at very low $Z'$
provided $\zeta/n$ is small enough for the hydrogen to
become molecular. (The time-scale for such conversion 
may be long at early cosmic times as we have estimated in \S \ref{sub: time scales}.)
The C/CO abundance ratio grows to $\sim 1$ at high $\zeta/n$
but not much larger, again so long as the conversion to H$_2$ 
is complete. 

Second, we find that the OH abundances are 
maximal near the critical H-to-H$_2$ transition points. This is
a very general trend also encapsulated in our analytic
expressions for the OH abundance in the atomic
and molecular regimes (Equations [\ref{eq: OH highZ}] and [\ref{eq: OH lowZ Z and zetan dependence}]). A large fraction of the available gas-phase oxygen is driven to OH
at low $Z'$, and the OH persists within the atomic regime
(e.g., Figure \ref{fig: 2d_OH_H2O_UV0}). To first-order, the OH abundance is
proportional to $\zeta/n$ in the molecular regime,
but inversely proportional to $\zeta/n$ in the atomic regime.

Third, the OH/CO abundance ratio approaches unity near 
the H-to-H$_2$ transition points, and increases without 
limit, as $1/Z'$, into the (low-metallicity) atomic regime.
The behavior of the OH/CO ratio is not affected by FUV photodissociation of the OH
so long as the CO is shielded by H$_2$.

Fourth, the O$_2$/OH ratio becomes large within the molecular regime at sufficiently low \z and increasing metallicity. In the molecular regime C/CO and O$_2$/OH are anticorrelated.
However, OH is always the dominant oxygen-bearing molecule at sufficiently low $Z'$. Furthermore, even for comparable carbon and oxygen elemental abundances, OH/CH $\gg 1$ at low metallicities, and OH becomes the dominant heavy-bearing molecule at low $Z'$.

Our study indicates that for ionization parameters characteristic of the Milky Way
and present-day galaxies,
much of the cold dense low-metallicity ISM for the Pop-II generation would have been OH-dominated and atomic rather than CO-dominated and molecular. However, for sufficiently low ionization parameters conversion to H$_2$ will occur for FUV-shielded gas. Because the gas phase processes enable complete
incorporation of carbon into CO even at very low metallicity, the CO abundance will remain linearly proportional to the overall metallicity and carbon abundance in this regime. Observationally, CO rotational line emissions then remain good proxy tracers for the H$_2$, but the emissions may become optically thin even for the dominant isotope, in which case the elemental carbon abundances affect 
the CO-to-H$_2$ conversion factors.  For high metallicity systems for which the H$_2$ and CO formation times are shorter, conversion to CO may 
nevertheless not always be complete when the ionization parameters 
are large even for fully FUV-shielded H$_2$ gas.  This is relevant for
observations of galaxies with elevated star-formation and enhanced cosmic-ray ionization rates.  For such systems, CO may not be a reliable tracer of 
the H$_2$ even in cosmic-ray or X-ray dominated cores
\citep[see also][]{Bisbas2015}.
Detailed radiative transfer computations are required to estimate CO emission
line brightness temperatures and associated CO-to-H$_2$ conversion factors
in this regime.

As we have demonstrated in this paper, large OH abundances may
persist in shielded atomic gas even at very low metallicities.
This is of particular interest because observations of (radio-wave)
``conjugate" OH rotational satellite lines may be used to probe
or set limits on 
spatial and temporal variations of fundamental quantities such as
the fine-structure-constant and/or the electron-to-proton mass ratio
\citep[e.g.,][]{Kanekar2004, Kanekar2012, Kozlov2013}.
Observations of OH at high redshift, perhaps feasible given the
large expected abundances, could provide a broad
cosmological baseline for investigating such variations.  
Probes via CH, e.g.~via the $\Lambda$-doublet transitions
\citep{deNijs2012, Truppe2013} would be much less promising.

In future searches for molecules at the reionization epoch
\citep[e.g.,][]{Carilli2011, Heywood2011, Lidz2011, Munoz2013} OH may be a better target than CO.

\section*{Acknowledgments}

We thank Alex Dalgarno, Avi Loeb, Evelyne Roueff, and Ewine van Dishoeck for helpful conversations during the course of this work. We thank the referees for helpful comments that have improved our paper.
S.B.~acknowledges support from the Raymond and Beverly Sackler Tel Aviv University -- Harvard Astronomy Program.
This work was also supported in part by the DFG via German -- Israeli 
Project Cooperation grant STE1869/1-1/GE625/15-1, and by a
PBC Israel Science Foundation I-CORE Program grant 1829/12.


\end{document}